\documentclass[pra,twocolumn,reprint]{revtex4-1}
\usepackage{amsmath,amssymb,amsthm,bm,braket,ascmac,bbm,color}
\usepackage{graphicx}
\usepackage{dsfont}

\newcommand{\dd}{\mathrm{d}}
\newcommand{\ee}{\mathrm{e}}
\newcommand{\ii}{\mathrm{i}}
\newcommand{\bfig}{\begin{figure}\begin{center}}
\newcommand{\efig}{\end{center}\end{figure}}

\newenvironment{sumipmatrix}{\left ( \begin{smallmatrix}} {\end{smallmatrix}\right )}

\newcommand{\spinor}{\phi}
\newcommand{\wfv}{\vec{\psi}}
\newcommand{\wf}{\psi}
\newcommand{\chiv}{\vec{\chi}}
\newcommand{\nuv}{\vec{\nu}}
\newcommand{\lamv}{\vec{\lambda}}

\newcommand{\fechi}{\bm{\chi}}
\newcommand{\fechinew}{\bm{\chi}'}
\newcommand{\fenu}{\bm{\nu}}
\newcommand{\felam}{\bm{\lambda}}

\newcommand{\hop}{\hat{\mathcal{H}}}

\newcommand{\hmat}{H}

\newcommand{\hfmat}{H^{\text{F}}}
\newcommand{\hftmat}{H^{\text{T,F}}}
\newcommand{\pert}{V^\text{F}}

\newcommand{\coeffF}{\mathcal{B}}
\newcommand{\coeff}{\mathcal{C}}
\newcommand{\eorg}{\epsilon}
\newcommand{\efield}{E}
\newcommand{\flcoeff}{w}

\newcommand{\idtwo}{\mathds{1}}

\newcommand{\hhd}{p}
\newcommand{\enhance}{\xi}
\newcommand{\res}{\ell}
\newcommand{\hres}{H^\text{F,res}}
\newcommand{\resmom}{k_\res}

\newcommand{\ob}{\Omega_\text{B}}
\newcommand{\nb}{N_\text{B}}
\newcommand{\nel}{N_{\mathrm{el}}}

\newcommand{\id}{\idtwo}

\newcommand{\curcoeff}{v}
\newcommand{\velres}{u}

\newcommand{\re}{\text{Re}}
\newcommand{\sxe}{\vec{\zeta}}
\newcommand{\avz}{\overline{|z_k|}}

\newcommand{\cur}{I}
\newcommand{\curH}{I_\text{H}}
\newcommand{\curC}{I_\text{C}}
\newcommand{\curHpre}{I^\alpha_\text{H}{}'}
\newcommand{\pow}{P_\text{rad}}
\newcommand{\jop}{\hat{\mathcal{I}}}
\newcommand{\jmat}{I}

\newcommand{\hhc}[1]{I_\text{H}(#1)}

\newcommand{\cre}[1]{\hat{#1}^\dag}
\newcommand{\ann}[1]{\hat{#1}}

\newcommand{\tc}{T_\text{c}}

\begin{document}
\title{
Floquet-Theoretical Formulation and Analysis of 
High-Harmonic Generation in Solids}

\author{Tatsuhiko N. Ikeda}
\affiliation{Institute for Solid State Physics, University of Tokyo, Kashiwa, Chiba 277-8581, Japan}
\author{Koki Chinzei}
\affiliation{Institute for Solid State Physics, University of Tokyo, Kashiwa, Chiba 277-8581, Japan}
\author{Hirokazu Tsunetsugu}
\affiliation{Institute for Solid State Physics, University of Tokyo, Kashiwa, Chiba 277-8581, Japan}

\date{\today}
\begin{abstract}
By using the Floquet eigenstates,
we derive a formula to calculate the high-harmonic components
of the electric current (HHC) in the setup where a monochromatic laser field is turned on at some time.
On the basis of this formulation, we study the HHC spectrum of electrons on a one-dimensional chain
with the staggered potential to study the effect of multiple sites in the unit cell such as the systems with charge density wave (CDW) order.
With the help of the solution for the Floquet eigenstates,
we analytically show that two plateaus of different origins emerge in the HHC spectrum.
The widths of these plateaus are both proportional to the field amplitude,
but inversely proportional to the laser frequency and its square, respectively.
We also show numerically that multi-step plateaus appear when both the field amplitude
and the staggered potential are strong.
\end{abstract}
\maketitle

\section{Introduction}
High-harmonic generation (HHG)
is the basis for the attosecond physics
and has attracted renewed attention
owing to its successful observations in bulk solids
driven by a strong laser field~\cite{Ghimire2011,Schubert2014,Hohenleutner2015,Luu2015,Ndabashimiye2016,Ghimire2014,Vampa2015,Yoshikawa2017,Kaneshima2018}.
These observations have revealed that
the HHG in solids has characteristics different from
those in atomic gases~\cite{Brabec2000}.
For example, the high-energy cutoff of the output spectrum scales linearly with
the input field amplitude~\cite{Ghimire2011,Schubert2014,Luu2015} rather than its square i.e.\ the laser intensity.
Besides, multiple plateaus emerge in the high-harmonic output spectrum for very large amplitude~\cite{Ndabashimiye2016}.
To understand the microscopic mechanism of these unique features of HHG in solids,
many theoretical and experimental studies are actively being conducted.

A theoretical approach to this problem
is to analyze the electron dynamics in solids in the time domain.
In this approach, the two-band semiconductor Bloch equation~\cite{Golde2006,Golde2008,Golde2011,Vampa2014,Tamaya2016a,Tamaya2016b}
and the time-dependent Schr\"{o}dinger equation~\cite{Plaja1992,Korbman2013,Wu2015,Du2017,Ikemachi2017,Jia2017,Ikemachi2017a}
were numerically solved in the presence of a pulse electric field,
and analysis was performed for
the high-harmonic components of the electric current (HHC), which work as the source of the HHG.
For both equations,
the unique scaling in solids is reproduced
in one-dimensional models
and 
it has been shown that 
the interband transition plays an important role as well as the intraband dynamics.
Thus the time-domain approach has successfully reproduced the experimental observations,
but it does not fit analytical approaches and
it is not straightforward to obtain systematic understanding of microscopic physics.
For instance, 
it is obscure why the output spectrum has peaks at multiples of the input laser frequency
since the pulse input has a continuous spectrum.

A complementary theoretical approach is to invoke the Floquet theory~\cite{Shirley1965}
and analyze the electron dynamics in the frequency domain.
In this approach, the input electric field is idealized to have an exact periodicity in time,
and this periodicity is utilized to define the Floquet eigenstates~\cite{Tzoar1975},
which correspond to the solutions of the time evolution equation.
For the time-dependent Schr\"{o}dinger equation, 
early studies~\cite{Faisal1989,Tal1993,Faisal1996,Faisal1997} analyzed the Floquet eigenstates and the HHC carried by them.
More recently, the Floquet theory has been applied
to one-dimensional systems~\cite{Martinez2002,Yan2008,Park2014,Ikeda2018},
graphene and carbon nanotubes~\cite{Gupta2003,Alon2004,Hsu2006},
and three-dimensional systems~\cite{Faisal2005}.
However, it has not been discussed well how the Floquet eigenstates
are related to the initial states in recent experiments.
In addition, those previous studies are mostly numerical,
and the characteristics of the HHG in solids have not been fully understood.
There are also Floquet-theoretical approaches for the semiconductor equation~\cite{Higuchi2014,Dimitrovski2017}.
Higuchi et al.~\cite{Higuchi2014} 
considered the quasistatic limit of the input electric field and
discussed the HHC originating from the Bloch oscillation.
The high harmonics induced by this mechanism are multiples of the Bloch frequency $\ob$,
not the input laser frequency $\Omega$,
and this regime differs from that of the experiments in Ref.~\cite{Ghimire2011,Ndabashimiye2016}.

In this paper, we develop a theoretical framework
based on the Floquet theory to investigate the mechanism of the HHG in solids.
Considering the setup that an ac electric field with frequency $\Omega$ is turned on at some time to drive the system,
we derive a formula to obtain the HHC spectrum from the Floquet eigenstates [see Eqs.~\eqref{eq:defhhc} and \eqref{eq:hhceigen}].
On the basis of our formulation,
we then analyze the HHC spectrum of
electrons on a one-dimensional chain [see Eq.~\eqref{eq:RMmodel}]
to study the effect of multiple sites in the unit cell.
We realize a two-site unit cell by introducing a staggered potential,
which changes its sign alternately along the chain.
In the absence of the staggered potential,
we see the presence of a plateau in the HHC spectrum for strong field.
Then we analytically show that the staggered potential induces another wider plateau,
which sets in already for a weaker field.
We argue that the widths of both plateaus scale linearly with the field amplitude
consistently with experimental observations.
A new prediction of our analysis is that,
for a fixed field amplitude,
the widths of the two plateaus scale as $\Omega^{-1}$ and $\Omega^{-2}$.
We then numerically calculate the HHC spectrum
and show that multi-step plateaus emerge when
both the field amplitude and the staggered potential are strong enough.

The rest of this paper is organized as follows.
In Sec.~\ref{sec:formulation},
we introduce our model Hamiltonian in the Floquet formulation.
We also summarize the properties of the Floquet eigenstates and explain
how to calculate the HHC spectrum.
Section~\ref{sec:symmetry}
summarizes the symmetry properties of the HHC spectrum.
In Sec.~\ref{sec:single},
we obtain an analytic form of the HHC spectrum
in the single-band limit, where the staggered potential is absent,
and discuss the plateau in the spectrum.
We also obtain the asymptotically exact Floquet eigenstates analytically for small or moderate field amplitude.
By using these analytic forms of eigenstates,
we develop
in Sec.~\ref{sec:pert}
a perturbation theory with respect to the staggered potential,
and discuss the new plateau induced by the potential.
In Sec.~\ref{sec:beyond},
we numerically analyze the HHC spectrum where the perturbation theory is not applicable.
Section~\ref{sec:conclusions}
summarizes the results with concluding remarks.
In the Appendix, we provide supplemental technical details
consolidating the discussions in the main text.

\section{Formulation of the problem}\label{sec:formulation}
In this section,
we derive formulas for the HHC
in terms of the Floquet eigenstates.
We investigate a situation in which
a monochromatic ac electric field is turned on at time $t=0$.
We solve the time-dependent Schr\"{o}dinger equation
by invoking the Floquet eigenstates and
calculate the Fourier components of the electric current for the solution.

\subsection{Model}
In this paper, we study the response of electron systems on a lattice
with unit cell containing multiple sites.
As the simplest model,
we consider a model of electrons on a one-dimensional chain
with the staggered potential, which doubles the size of the unit cell.
We note that it is straightforward to generalize the following arguments
to the cases of potentials with periodicity larger than two and the results do not change qualitatively.
The Hamiltonian is given by 
\begin{equation}\label{eq:RMmodel}
    \hop_0=\sum_{j=1}^{2L} \left[ t_0(\cre{c}_j \ann{c}_{j+1}+\cre{c}_{j+1} \ann{c}_j)
    +Q(-1)^j \cre{c}_j \ann{c}_j \right],
\end{equation}
where $\ann{c}_j$ ($\cre{c}_j$) is the annihilation (creation) operator
for the electron on the $j$-th site.
We have ignored the spin degrees of freedom,
which, if considered, only multiply the following results by the factor of 2.
The length of the chain is given by $2L$ with an even number $L$ $(>0)$,
and the periodic boundary condition is imposed.
The parameter $t_0$ denotes
the transfer integral,
$Q$ is the amplitude of the staggered potential,
and we restrict ourselves to the case of $|Q|<1$ in this paper.

This model is often used to study the interplay between the Bloch oscillation
and the Zener tunneling~\cite{Breid2006,Mizumoto2012,Mizumoto2013} since the staggered potential $Q$ splits the single cosine band into two.
The physical systems described well by this model include
some binary compounds with chemical formula AB
and the electrons in the presence of static CDW order with period two.
In the following, 
the unit of energy is fixed so that $t_0=1/2$.
This implies that the half of the total band width for $Q=0$ is set to unity in our unit.
Thus our unit of energy is read typically as 2.5\,eV for semiconductors
and 0.25\,eV for one-dimensional organic conductors.

We make a remark on the relationship between our model~\eqref{eq:RMmodel}
and the two-band models for typical semiconductors.
Generally speaking, multiple bands are formed from several single-electron states in
the unit cell, and there are two typical cases.
In the first case, the band multiplicity corresponds to the number of
different atomic orbitals at each site, and this is a standard setup for semiconductors.
In the second case, the band multiplicity is the number of different
sublattice sites in the unit cell, and we focus on this case in this paper.
A tight-binding Hamiltonian can model both cases,
and applying the electric field generally induces interband transitions of electrons
regardless of the origin of multiple bands,
although their matrix elements depend on details such as the type of atomic orbitals and the position of sublattice sites.

The Hamiltonian~\eqref{eq:RMmodel}
is diagonalized in the momentum space
by the Fourier transformation for two-site unit cells:
$\ann{a}_k = L^{-1/2}\sum_{j=1}^L \ee^{-\ii k(2j)} \ann{c}_{2j}$
and $\ann{b}_k = L^{-1/2}\sum_{j=1}^L \ee^{-\ii k(2j+1)} \ann{c}_{2j+1}$.
Here
the distance $a$ between the neighboring sites is set to unity,
and the lattice momentum $k$ takes the values of $k=\pi n/L$ $(n=-L/2,-L/2+1,\dots,L/2-1)$.
By substituting these Fourier transforms into Eq.~\eqref{eq:RMmodel},
we obtain
\begin{align}
\hop_0  &=\sum_k
   \cre{\spinor}_k 
    \hmat_0 (k)
    \ann{\spinor}_k \label{eq:H0k}
\end{align}
with $\cre{\spinor}_k = (\cre{a}_k \ \cre{b}_k)$
and the $2\times2$ Hamiltonian matrix
\begin{align}\label{eq:h0mat}
\hmat_0(k)=\cos k \, \sigma_x +Q\sigma_z,
\end{align}
where $\sigma_x$ and $\sigma_z$ are the Pauli matrices.
The two eigenvalues of $\hmat_0(k)$ are given by
\begin{align}\label{eq:eorg}
\eorg_\pm(k)  = \pm\sqrt{\cos^2 k+Q^2}
\end{align}
and we refer to $\eorg_+(k)$ and $\eorg_-(k)$
as the upper and the lower bands, respectively.
The band gap is given by $2|Q|$, which is the energy 
difference at the Brillouin-zone boundary $k=-\pi/2$~\footnote{
The energy gap minimum can be moved to $k=0$ by a unitary transformation $\ann{c}_j\to \ann{c}_j \exp(\ii\frac{\pi}{2}\frac{j}{2})$,
which shifts the electron momentum $k\to k+\pi/2$}.

The effects of the time-dependent electric field $\efield(t)$
are taken into account in terms of the vector potential $A(t)$,
which satisfies $\dd A(t)/\dd t=-E(t)$.
Throughout this paper,
we assume that the electric field and, hence, the vector potential
are homogeneous in space.
The vector potential 
modifies the Hamiltonian~\eqref{eq:RMmodel} by the gauge-invariant Peierls substitution:
$\cre{c}_{j_1}\ann{c}_{j_2} \to \cre{c}_{j_1} \ann{c}_{j_2} \ee^{\ii eA(t)(j_1-j_2)}$,
where $-e$ denotes the electron charge.
Correspondingly,
Eq.~\eqref{eq:H0k} is replaced by
\begin{align}\label{eq:Hk}
\hop(t)  &=\sum_k
    \cre{\spinor}_k
    \hmat(k,t)
    \ann{\spinor}_k
\end{align}
where
\begin{align}
\hmat(k,t)= \hmat_0 (k+eA(t)) = \cos\left[ k+eA(t)\right]\sigma_x+Q\sigma_z.
\end{align}
We note that this time-dependent Hamiltonian is diagonal in $k$
since the vector potential does not break the translation symmetry.

\subsection{Time evolution}\label{sec:evolution}
In the present work, we consider the dynamics induced by
the ac electric field $E(t)$ with frequency $\Omega$ $(>0)$,
which is turned on at time $t=0$:
$E(t)=-E_0\cos \Omega t$ for $t>0$ and $E(t<0)=0$.
This is represented by the following vector potential:
\begin{align}\label{eq:Adef}
A(t) = A_0 \theta(t) \sin \Omega t; \qquad A_0=\frac{E_0}{\Omega}.
\end{align}
The time dependence in the Hamiltonian~\eqref{eq:Hk} is now given by
\begin{align}\label{eq:Hkmat}
\hmat(k,t) = \cos( k +F\sin \Omega t) \sigma_x +Q\sigma_z
\end{align}
for $t>0$.
Here the dimensionless parameter
\begin{align}\label{eq:Fdef}
F\equiv eA_0 = \frac{eE_0}{\Omega}=\frac{\Omega_\text{B}}{\Omega}
\end{align}
quantifies the strength of the coupling to the input electric field
and $\Omega_\text{B} =eE_0$ is the so-called Bloch frequency.

Our monochromatic input~\eqref{eq:Adef} has two advantages.
First, the high harmonics
are well defined as multiples of $\Omega$,
in contrast to polychromatic inputs such as a pulse~\footnote{
Since the HHG is a nonlinear response and the superposition principle does not hold,
the results for the monochromatic and the polychromatic inputs
are not related simply to each other.}.
Second,
the time-dependent Hamiltonian $H(k,t)$~\eqref{eq:Hkmat} becomes periodic in $t>0$ with period $T=2\pi/\Omega$.
We will utilize this periodicity in the following to solve the time-dependent Sch\"{o}dinger equation.
We note that the input~\eqref{eq:Adef} has additional symmetries $A(t)=-A(T/2+t)=-A(T-t)$, 
which imply $\hmat(-k,t)=\hmat(k,t+T/2)=H(k,T-t)$.

As for the initial condition ($t<0$),
we consider the case that
the electron density is half-filling, $\langle \cre{c}_j \ann{c}_j\rangle+\langle \cre{c}_{j+1} \ann{c}_{j+1}\rangle=1$ in each unit cell,
and the system is in the ground state of $\hop_0$:
\begin{align}
\ket{\Psi_0}=\prod_k \left[\cre{\spinor}_k \cdot \wfv^0_k \right]\ket{0};
\quad \wfv^0_k =\begin{pmatrix} \wf^0_{k,a} \\ \wf^0_{k,b} \end{pmatrix},
\end{align}
where the product runs over all $k$'s in the Brillouin zone
and $\ket{0}$ denotes the Fock vacuum~\footnote{
The arguments in Sec.~\ref{sec:formulation}
can apply to any initial state in the form of a single Slater determinant
$\ket{\Psi_0}=\prod_{j=1}^{\nel} \left[\cre{\spinor}_{k_j} \cdot \wfv^0_{k_j} \right]\ket{0}$,
where $\nel$ denotes the number of electrons
and their momenta are $\{ k_1,k_2,\dots,k_{\nel} \}$.
}.
Namely, $\wfv^0_k$ corresponds to 
the one-particle wave function with the negative energy $\eorg_-(k)$.
Then we are interested in the evolution of the many-body state $\ket{\Psi(t)}=\prod_k  [\cre{\spinor}_k \cdot \wfv_k(t) ]\ket{0}$,
in which each $\wfv_k(t)$ obeys the one-particle Schr\"{o}dinger equation
\begin{align}\label{eq:tdSch}
\ii \frac{\dd}{\dd t} \wfv_k(t) = \hmat (k,t) \wfv_k(t)
\end{align}
with the initial condition $\wfv_k(t=0)=\wfv_k^0$.
Here we have used the fact that
$\hop(t)$ is diagonal in $k$
and 
the particle number $\cre{\spinor}_k \ann{\spinor}_k$ for each $k$ is conserved.

Owing to the periodicity in time,
the general solutions of the time-dependent Sch\"{o}dinger equation~\eqref{eq:tdSch}
are obtained by the Floquet theory~\cite{Shirley1965}.
The Floquet Hamiltonian is given by~\footnote{We implicitly assume $A(t)=A_0\sin\Omega t$ for $-\infty< t<\infty$ in using the Floquet theory.
Once we impose the initial condition $\wfv_k(0)=\wfv_k^0$, the solution of Eq.~\eqref{eq:tdSch} is appropriately obtained in $t\ge0$.
}
\begin{align}\label{eq:HF}
\hfmat_{mn}(k) &= n\Omega \delta_{mn}\idtwo+ \int_0^T\frac{\dd t}{T} \hmat(k,t) \ee^{-\ii (m-n)\Omega t} \\
&= (n\Omega \idtwo+Q\sigma_z)\delta_{mn}\notag\\
&\qquad+ \ii^{m-n}J_{m-n}(F)\cos\left[ k-\frac{(m-n)\pi}{2}\right] \sigma_x
\end{align}
for each pair of integers $m$ and $n$ ($-\infty<m,n<\infty$).
Here
$\idtwo$ is the $2\times2$ unit matrix,
and $J_n(z)$ denotes the Bessel function of the first kind.
The Floquet eigenstates $\fechi(k) =\{ \chiv_n(k)\}_n$ are defined by
\begin{align}
\sum_n \hfmat_{mn}(k) \chiv_n^\alpha(k) = E_\alpha(k) \chiv_m^\alpha(k),\label{eq:corrst}
\end{align}
and $E_\alpha(k)$'s are the Floquet eigenvalues.
The corresponding time-dependent physical state is given by
\begin{align}\label{eq:flsol}
\chiv^\alpha(k,t) =\ee^{-\ii E_\alpha(k)t} \sum_n \chiv_n^\alpha(k) \ee^{\ii n\Omega t}.
\end{align}
This becomes a solution of Eq.~\eqref{eq:tdSch}
and satisfies $\chiv^\alpha(k,t+T)=\ee^{-\ii E_\alpha(k)T}\chiv^\alpha(k,t)$.

Although the Floquet Hamiltonian~\eqref{eq:HF} has an infinite number of eigenstates,
only two of them are physically independent
and this number is the dimension of $\hmat(k,t)$.
This is because, if $\{\chiv_n(k)\}_n$ is a Floquet eigenstate with eigenvalue $E(k)$,
then, shifting this in the Floquet space by any integer $M$ leads to another eigenstate
$\{\chiv_{n+M}(k)\}_n$ with eigenvalue $E(k)-M\Omega$,
and these shifted eigenstates all describe the same evolution~\eqref{eq:flsol}.
To avoid this redundancy, we may choose the two Floquet eigenstates with eigenvalues
in the interval $[-\Omega/2,\Omega/2)$, for example,
then they are always inequivalent.
In the following, we let $\fechi^\alpha(k)=\{\chiv_n^\alpha(k)\}_n$ [$\alpha=1,2$ and $E_1(k)\le E_2(k)$]
be the inequivalent Floquet eigenstates thus obtained,
which are normalized and orthogonal to each other $\fechi^\alpha(k)^\dag \fechi^\beta(k)=\delta^{\alpha\beta}$.

Equation~\eqref{eq:tdSch} with our initial condition
is solved by expanding the initial state $\wfv_k^0$
in terms of $\chiv^\alpha(k,t=0)=\sum_n \chiv^\alpha_n(k)\equiv \vec{X}^\alpha(k)$.
This expansion is always possible since $\sum_{\alpha}\vec{X}^\alpha(k)\vec{X}^{\alpha\dag}(k)=\id$,
and the expansion coefficients are calculated as $\flcoeff_\alpha(k)=\vec{X}^\alpha(k)^\dag \wfv^0_k$,
which satisfy $\sum_{\alpha}|\flcoeff_\alpha(k)|^2=1$.
Then, the solution of Eq.~\eqref{eq:tdSch} is given by
\begin{align}
\wfv_k(t)
= \sum_{\alpha} \flcoeff_\alpha(k)  \ee^{-\ii E_\alpha(k)t} \sum_n \chiv^\alpha_n(k) \ee^{\ii n\Omega t}.\label{eq:flsol2}
\end{align}
We note that Eq.~\eqref{eq:flsol2} holds true only for $t\ge0$,
and $\wfv_k(t<0)=\ee^{-\ii \eorg_-(k)t}\wfv_k^0$ for $t\le0$.

\bfig
\includegraphics[width=8cm]{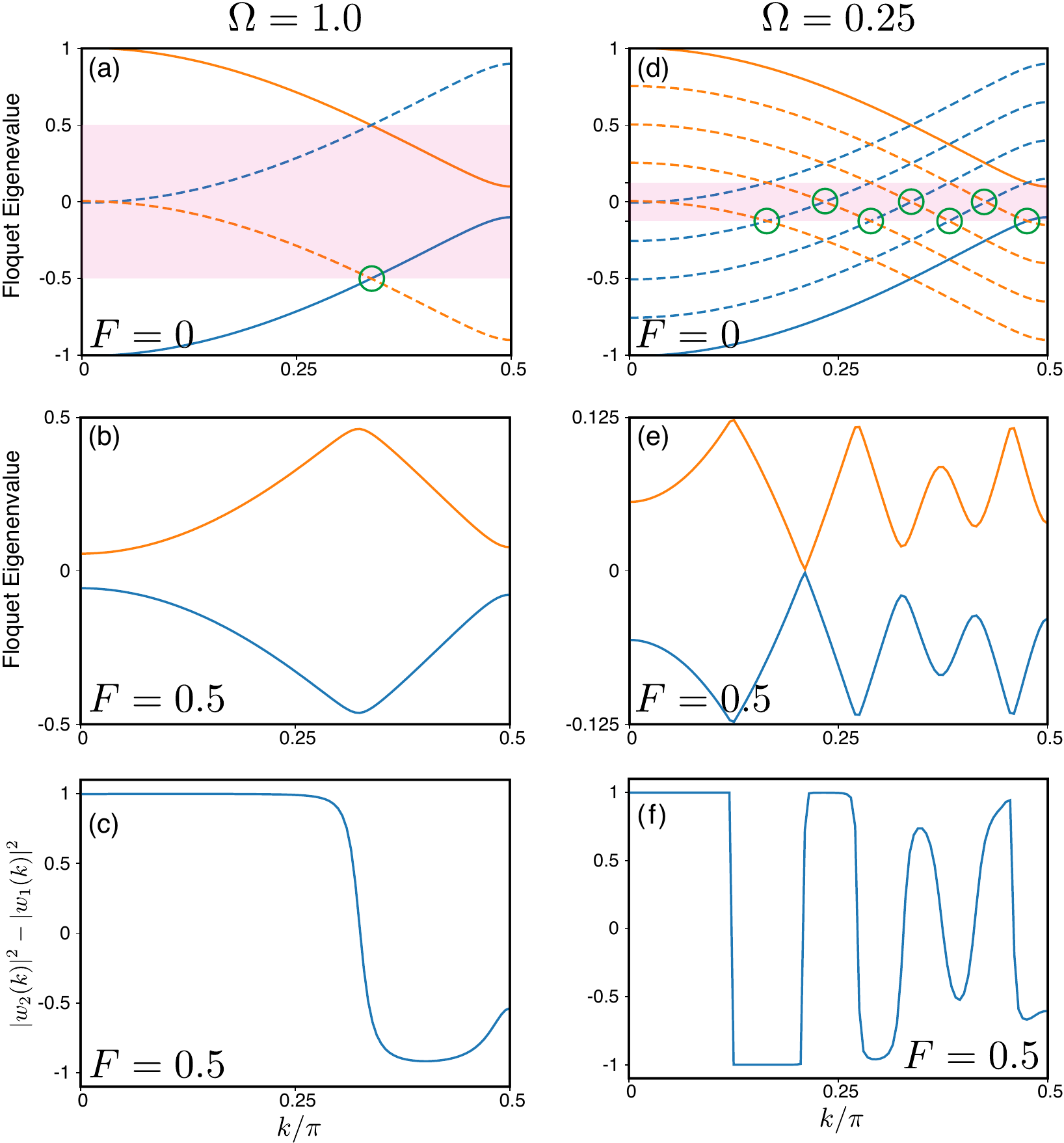}
\caption{
(a) Floquet eigenvalues for $F=0$ and $\Omega=1.0$.
The solid lines correspond to the energy bands $\eorg_\pm(k)$ [Eq.~\eqref{eq:eorg}],
and the dashed lines to the other Floquet bands.
The shaded area represents the representative interval $[-\Omega/2,\Omega/2)$,
in which the point of band crossing is encircled.
For $F=0.5$, (b) the representative Floquet eigenvalues
and (c) the weight difference $ |\flcoeff_2(k)|^2- |\flcoeff_1(k)|^2$
are plotted against $k$.
The panels (d)-(f) are similar plots to (a)-(c),
where $\Omega$ is changed to $0.25$.
In all panels, we have set $Q=0.1$.
}
\label{fig:disp}
\efig

Numerically,
the Floquet eigenstates
are obtained by diagonalizing the Floquet Hamiltonian $\hfmat_{mn}(k)$
with a sufficiently large cutoff for $|m|$ and $|n|$,
and the expansion coefficients $\flcoeff_\alpha (k)$ are calculated from them.
Figure~\ref{fig:disp}
shows in the left column the results for $\Omega=1$.
First, the panel (a) shows the Floquet eigenvalues for $F=0$ for reference, i.e.\
$\eorg_{\pm}(k)-M \Omega$ for some $M$'s.
The panel (b) shows the representative
Floquet eigenvalues $E^\alpha (k)$ for $F=1$.
They change smoothly with $F$ at most $k$'s,
but the crossing points of the different bands at $F=0$ become anticrossings at $F\neq0$.
The panel (c) shows the weight difference  $ |\flcoeff_2(k)|^2- |\flcoeff_1(k)|^2$
of the two Floquet eigenstates.
It changes sign at the anticrossing points, and its modulus reduces
also at the Brillouin-zone boundary.
Figures~\ref{fig:disp}(d)-(f) show the corresponding data for $\Omega=0.25$.
The lower input frequency increases the number of anticrossing points in the Floquet bands,
and this results in more oscillations in the weight difference.

\subsection{High-harmonic current (HHC)}
Here we use the solution~\eqref{eq:flsol2} represented by the Floquet eigenstates~\eqref{eq:corrst} to
calculate the time evolution of the electric current, which is the source of radiation.
We will show that the current spectrum consists of 
a discrete part peaked at $n\Omega$ $(n\in\mathbb{Z})$
and a continuous part.
The former, the high-harmonic current, works as the source of HHG. 

The electric current density is obtained as the expectation value
of the operator
\begin{align}\label{eq:current_op}
\jop (t)\equiv \frac{1}{2L}\frac{\partial \hop(t)}{\partial A(t)} =\frac{1}{2L}\sum_k \cre{\spinor}_k \jmat_k(t) \ann{\spinor}_k,
\end{align}
which is again diagonal in $k$ as seen from Eq.~\eqref{eq:Hk}.
Since $A(t)$ is periodic in time, $\jmat_k(t)$ is also periodic
and, hence, expanded in a Fourier series as
\begin{align}\label{eq:current_mat}
\jmat_k(t) &=\frac{\partial \hmat(k,t)}{\partial A(t)}= -e \sin(k+F\sin\Omega t)\sigma_x\notag\\
&= \sum_n \jmat_{k,n} \ee^{-\ii n \Omega t}.
\end{align}
We note that $\jmat_{k,-n}=\jmat_{k,n}^\dag$ since $\jmat_k(t)$ is Hermitian.
The $2\times2$ matrix $\jmat_{k,n}$
is given by
\begin{align}\label{eq:jmat}
\jmat_{k,n}
&= -\ii^n eJ_n(F) \sin \left(k+\frac{n\pi}{2} \right) \sigma_x \equiv -e\curcoeff_n(k)\sigma_x.
\end{align}
Equation~\eqref{eq:jmat} follows from Eqs.~\eqref{eq:Hk} and \eqref{eq:h0mat}
and the hermiticity implies $\jmat_{k,-n}=(-1)^n \jmat_{k,n}=\jmat_{k,n}^*$.
Evaluating the expectation value of Eq.~\eqref{eq:current_op}
for the solution~\eqref{eq:flsol2},
we obtain
\begin{align}\label{eq:current_exp}
\cur(t)&= \braket{\Psi(t)| \jop(t)| \Psi(t)}= \frac{1}{2L} \sum_k \wfv_k(t)^\dag \jmat_k (t) \wfv_k(k)\\
&=\frac{1}{2L}\sum_{k,\alpha,\beta}
\flcoeff_\alpha(k)^* \flcoeff_\beta(k)
\sum_{m,n,l}
\chiv^\alpha_m(k)^\dag \jmat_{k,l}  \chiv^\beta_n(k)
\notag\\ &\qquad\qquad\qquad\times \ee^{-\ii[ E_\beta(k)-E_\alpha(k) +(m+l-n)\Omega]t}.
\end{align}

Equation~\eqref{eq:current_exp} consists of two kinds of contributions,
which are diagonal ($\alpha=\beta$) and off-diagonal ($\alpha\neq\beta$)
in terms of the labels for the Floquet eigenstates~\footnote{
In calculating the Fourier components,
one should note that Eq.~\eqref{eq:current_exp} holds only for $t\ge0$ and $I(t<0)=0$.
}:
\begin{align}
I(t) &= \curH(t) + \curC(t),\\
\curH(t)&\equiv \sum_N \curH(N)\ee^{-\ii N\Omega t},\\ 
\curC(t) &\equiv \int_{-\infty}^\infty \frac{\dd \omega}{2\pi} \curC(\omega)\ee^{-\ii\omega t}.
\end{align}
The part $\curH(t)$ corresponds to the diagonal part, and we have
\begin{align}
&\hhc{N} \equiv \frac{1}{2L}\sum_{k,\alpha} |\flcoeff_\alpha(k)|^2 \curH^\alpha(k,N),\label{eq:defhhc} \\
&\cur_\text{H}^\alpha(k,N) \equiv \sum_{n,l}\chiv^\alpha_{n-l+N}(k)^\dag \jmat_{k,l} \chiv^\alpha_n(k).\label{eq:hhceigen}
\end{align}
This contribution only involves a discrete set of the harmonics of the input frequency $\Omega$.
On the other hand,
$\curC(t)$ involves the $k$-dependent frequencies $E_\alpha(k)-E_\beta(k)$,
which lead to a continuous spectrum $\curC(\omega)$ in the thermodynamic limit, $L\to\infty$.
In the time domain, $\curC(t)$ decays~\footnote{
To prove this, we note the following sum rule: $I(0)=\sum_N \hhc{N}+\int_{-\infty}^\infty \frac{\dd \omega}{2\pi} \curC(\omega)$.
Here, $\int_{-\infty}^\infty \frac{\dd \omega}{2\pi} \curC(\omega)$ and, hence $\int_{-\infty}^\infty \frac{\dd \omega}{2\pi} |\curC(\omega)|^2$
are finite since both $I(0)$ and $\sum_N \hhc{N}$ are finite.
Thus it follows from the Parseval-Plancherel identity $\int_{-\infty}^\infty \dd t |\curC(t)|^2
= \int_{-\infty}^\infty \frac{\dd \omega}{2\pi} |\curC(\omega)|^2$
that $\lim_{t\to\infty}\curC(t)=0$
}
whereas $\hhc{t}$ remains to oscillate as $t\to\infty$,
and the harmonic part dominates for sufficiently large $t$.

Thus we regard Eq.~\eqref{eq:defhhc} as the source of the $N$-th order HHG
and refer to $\curH(N)$ as the HHC spectrum.
According to the classical electromagnetism,
the total radiation power $\pow(N)$ from the $N$-th HHC is proportional to $(N\Omega)^2 |\hhc{N}|^2$.
In the following, we investigate the symmetry aspects and the $N$-dependence of $\hhc{N}$
rather than $\pow(N)$ for comparison to the related theoretical studies.
One can immediately obtain $\pow(N)$ from $\hhc{N}$ if necessary,
and the plateaus discussed below will become clearer when plotted for $\pow(N)$
due to the factor $(N\Omega)^2$.

We note that Eq.~\eqref{eq:defhhc} is a generalization
of similar formulas in the literature (see e.g.\ Eq.~(28) in Ref.~\cite{Hsu2006}).
In the literature, it is assume that some of the Floquet bands are fully occupied 
and the quantity
\begin{align}\label{eq:curHpre}
\curHpre(N) =  \frac{1}{2L}\sum_{k} \curH^\alpha(k,N)
\end{align}
is discussed.
On the other hand, 
our formula~\eqref{eq:defhhc} involves the weight factor $|\flcoeff_\alpha(k)|^2$ on each Floquet eigenstate,
and $|\flcoeff_\alpha(k)|^2$ can take any value between 0 and 1.
The effects of the fractional weight factor become more significant for a stronger electric field
or near the Brillouin-zone boundary and the anticrossing points as shown in Fig.~\ref{fig:disp}.

\section{Symmetry Aspects}\label{sec:symmetry}
Formulas~\eqref{eq:defhhc} and \eqref{eq:hhceigen} for the HHC based on the Floquet eigenstates
have the advantage that the consequences of the symmetry are manifest.
In this section, we first reproduce the important known property that
$\curHpre(N)$~\eqref{eq:curHpre} vanishes for any even $N$
owing to the inversion symmetry (see e.g.\ Ref.~\cite{Faisal1997}).
Then we discuss the symmetry of the weights $|\flcoeff^\alpha(\pm k)|^2$
and show that $\cur_\text{H}(N)$~\eqref{eq:defhhc} vanishes for any even $N$ in our choice of the vector potential $A(t)$,
although this does not hold once the initial phase of the input filed is shifted.

We begin by noting that the inversion symmetry
\begin{align}\label{eq:invH0}
\hmat_0(-k) = \hmat_0(k)
\end{align}
breaks down at time $t>0$ in the presence of the electric field.
However, there exists another symmetry combined with half-period time translation $H(-k,t+T/2)=H(k,t)$ 
owing to $A(t+T/2)=-A(t)$.
This leads to the following symmetry for the Floquet Hamiltonian
\begin{align}
\hfmat_{mn}(-k)=(-1)^{m-n} \hfmat_{mn}(k),
\end{align}
which implies
$\chiv^\alpha_n(-k)=(-1)^n\chiv^\alpha_n(k)$
for an appropriate choice of the overall phase.
Together with $\jmat_{-k,n}=(-1)^{n+1}\jmat_{k,n}$,
it follows from Eq.~\eqref{eq:hhceigen}
\begin{align}\label{eq:Jinveigen}
\curH^\alpha(-k,N) = (-1)^{N+1} \curH^\alpha(k,N).
\end{align}
This means $\curHpre(N)=0$ for even $N$
since the contributions from $\pm k$ cancel out with each other.
We remark that Eq.~\eqref{eq:hhceigen} holds true
also for any inputs as long as $A(t+T/2)=-A(t)$ is satisfied.

The inversion symmetry between the weights $|\flcoeff^\alpha(\pm k)|^2$ 
follows from yet another symmetry $H(-k,T-t)=H(k,T+t)$, which leads to 
\begin{align}
\hfmat_{mn}(-k) = \hfmat_{nm}(k)= \hfmat_{mn}(k)^*.
\end{align}
This implies 
$\chiv^\alpha_n(-k)=\chiv^\alpha_n(k)^*$ and $\vec{X}^\alpha(-k)=\vec{X}^\alpha(k)^*$
with appropriate choices of the overall phases.
Besides, since Eqs.~\eqref{eq:invH0} and \eqref{eq:h0mat} ensure $\wfv_{-k}^0=\wfv_k^0=\wfv_k^{0}{}^*$,
we obtain $\flcoeff_\alpha(-k)=\flcoeff_\alpha(k)^*$ and, hence,
\begin{align}\label{eq:invw}
|\flcoeff_\alpha(-k)|^2=|\flcoeff_\alpha(k)|^2
\end{align}
for all $\alpha$'s.
The inversion symmetry of $|\flcoeff^\alpha(\pm k)|^2$~\eqref{eq:invw},
together with Eq.~\eqref{eq:Jinveigen},
means that the HHC spectrum~\eqref{eq:defhhc}
vanishes for even $N$'s.

We note that this property~\eqref{eq:invw} is violated
if we shift the initial phase of the input as $A(t)=A_0\sin (\Omega t+\theta_0)$.
When $0<\theta_0<\pi$,
the time evolution for $\pm k$ occur asymmetrically, 
and we have $\chiv^\alpha_n(-k)=\chiv^\alpha_n(k)^*\ee^{2\ii n\theta_0}$
and $|\flcoeff_\alpha(-k)|^2\neq|\flcoeff_\alpha(k)|^2$.
This asymmetry in the weights of Floquet eigenstates
leads to $\curH(N)\neq0$ for even $N$'s even if $\hop_0$ has the inversion symmetry.
In the following, we restrict ourselves to the case of $\theta_0=0$
and discuss the $N$-dependence of the HHC spectrum for odd $N$'s.

\section{single-band limit}\label{sec:single}
In this section, we discuss the case where $Q=0$
and the staggered potential is absent.
We refer to this case as the single-band limit
because we can also use the single-site unit cell and the Brillouin zone is doubled,
where each $k$ has only one energy band $\cos k$.
However, to compare with the case of $Q\neq0$,
we keep using the two-site unit cell.
As discussed in Ref.~\cite{Pronin1994},
the HHC are present due to the nonlinearity of the Peierls substitution,
in contrast to the continuous models (see e.g., Ref.~\cite{Ikeda2018}).
In Sec.~\ref{sec:intra},
we calculate the HHC spectrum for $Q=0$ in our formulation
and obtain results consistent with Ref.~\cite{Pronin1994}.
In Sec.~\ref{sec:fe_single},
we derive the asymptotically exact eigenstates for the Floquet Hamiltonian
for small $|F|$ $(\lesssim1)$,
which will serve as the basis for analyzing the $Q\neq0$ case
in Sec.~\ref{sec:pert}.

\subsection{The HHC spectrum}\label{sec:intra}
In this special case of $Q=0$, the time-evolution operator for Eq.~\eqref{eq:tdSch} is exactly obtained
since the Hamiltonians at different times commute with each other.
In the $2\times2$ matrix form, the time-evolution operator is given by
\begin{align}\label{eq:evolU}
U_k(t) =  \exp\left[ -\ii \sigma_x \int_0^t \dd \tau \, \cos( k +F \sin \Omega \tau)  \right].
\end{align}
It is noteworthy that this commutes with the current matrix [Eqs.~\eqref{eq:current_mat} and \eqref{eq:jmat}],
and therefore $U_k(t)^\dag \jmat_k(t) U_k(t)=\jmat_k(t)$.


It is straightforward to calculate the HHC spectrum 
from Eqs.~\eqref{eq:current_mat} and \eqref{eq:jmat}.
Noting that the initial state $\wfv_k^0$ is the eigenvector of $\sigma_x$ with $-1$ eigenvalue,
we obtain
\begin{align}\label{eq:Nth0k}
\hhc{k,N}=e\ii^N J_N(F)\sin\left(k+\frac{N\pi}{2}\right) 
\end{align}
Its $k$-sum vanishes 
for even $N$'s, and this is consistent with the inversion symmetry as discussed in Sec.~\ref{sec:symmetry}.
For odd $N$'s, we obtain
\begin{align}\label{eq:Nth0}
\curH (N) = \lim_{L\to\infty} \frac{1}{2L} \sum_k\curH (k,N) = \frac{\ii e}{\pi} J_N(F).
\end{align}
We note that, for an initial state at arbitrary filling,
the sum over $k$  in Eq.~\eqref{eq:Nth0} is restricted,
and the result is multiplied by $\sin( \pi \rho)$,
where $\rho$ is the electron density and $1/2$ at half filling.

We remark that the expectation value of the current
is also obtained at arbitrary time directly from Eq.~\eqref{eq:current_mat}
as
\begin{align}\label{eq:tdcurrent}
\cur(t) = \frac{e}{\pi}\sin\left( F\sin\Omega t\right)
\end{align}
in the limit of $L\to\infty$.
One can check that the Fourier expansion of Eq.~\eqref{eq:tdcurrent} reproduces Eq.~\eqref{eq:Nth0}.
Furthermore,
Eq.~\eqref{eq:tdcurrent} implies that the HHC spectrum contains only harmonics of $\Omega$
and the continuous part does not exist, $\curC(t)=0$, in the single-band limit.

Equation~\eqref{eq:Nth0} implies that the HHC spectrum qualitatively changes
depending on whether $|F|<1$ or $|F|>1$.
To understand this, we note that
$|J_N(F)|$ with fixed $F$ is approximately constant for $|N| \lesssim |F|$
and rapidly decays for $|N| \gtrsim |F|$.
Thus, when $|F|<1$, the HHC spectrum merely decays as $|N|$ increases.
On the other hand, when $|F|>1$,
a plateau emerges in the spectrum
and its width is given by $|F|$, which is proportional to $E_0$ and $\Omega^{-1}$.
These features are shown in Fig.~\ref{fig:intra}.
This plateau is essentially the same as the one discovered by Pronin and coworkers~\cite{Pronin1994}.

We note, however, that
this plateau is too narrow to explain the experiment by Ghimire et al.~\cite{Ghimire2011} that
detected the harmonics up to the 25th at $F\sim 5$.
They also showed the presence of a wider plateau if the band dispersion is deformed from $\cos k$
even in the single-band case.
Later in Sec.~\ref{sec:pert},
we will show another mechanism for a wider plateau, i.e.,
the staggered potential $Q$ splitting a single band into two.
This wider plateau has a different scaling of its width with input frequency $\Omega$.

\bfig
\includegraphics[width=6cm]{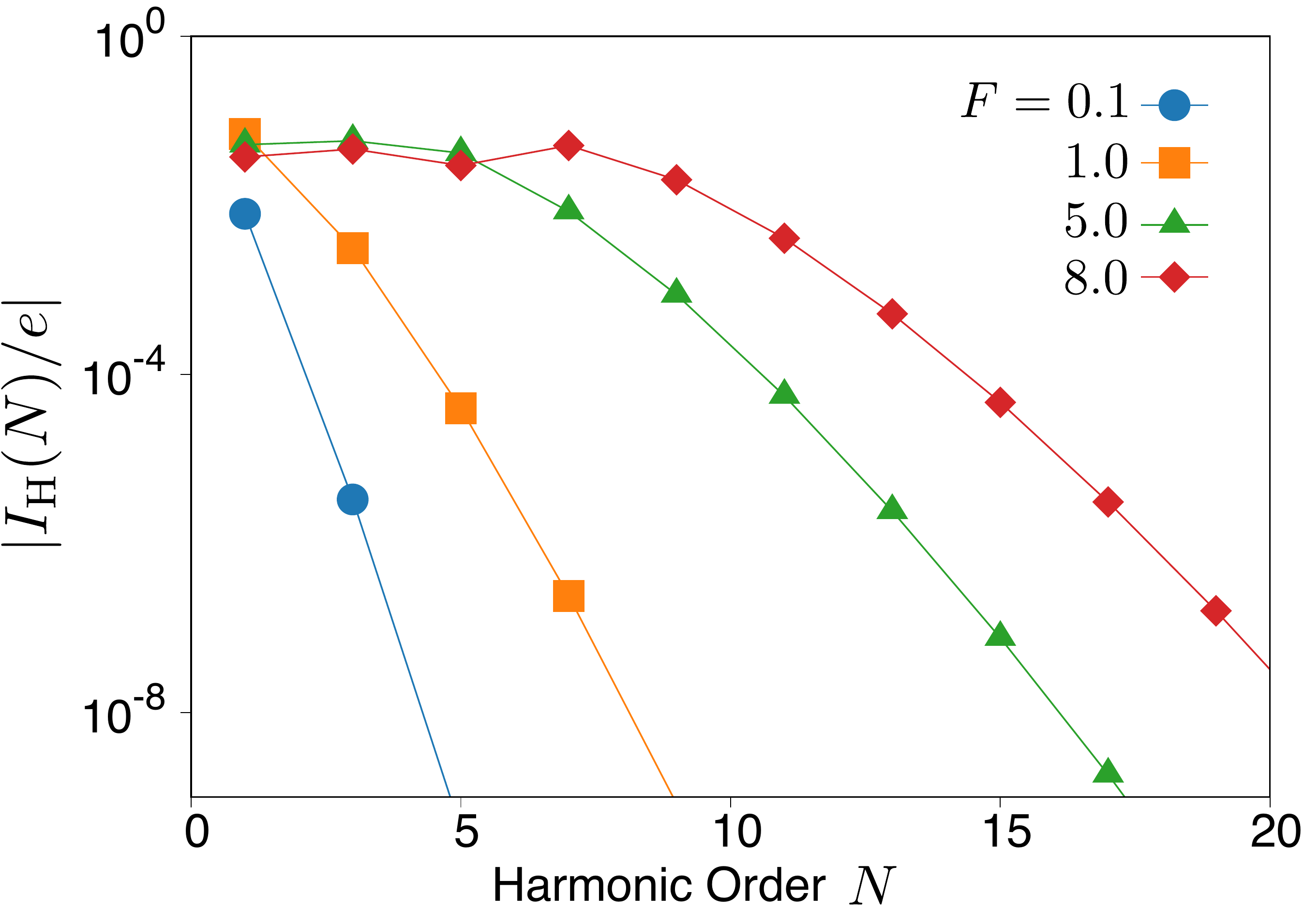}
\caption{The HHC spectrum in the single-band limit $Q=0$ [Eq.~\eqref{eq:Nth0}].
Each data set corresponds to the field strength $F=0.1$ (circle), 1.0 (square), 5.0 (triangle), and 8.0 (diamond).
}
\label{fig:intra}
\efig

\subsection{Floquet Eigenstates at $|F|<1$}\label{sec:fe_single}
We will show in Sec.~\ref{sec:pert} that
the staggered potential $Q$ produces another plateau even when $|F|<1$.
For this purpose, we here derive the Floquet eigenstates $\fechi^\alpha(k)$
in the single-band limit.

To obtain the Floquet eigenstates,
we first calculate the Fourier expansion of the time-evolution operator $U_k(t)$.
Since the exact result is very complicated,
we consider the case of $|F|<1$
and approximate the expansion
$\cos(k+F\sin \Omega t)=\sum_n  J_n(F) \re[\ee^{\ii (k+n\Omega t)}]$
in Eq.~\eqref{eq:evolU}
by its partial sum of $-1\le n\le 1$.
Within this approximation,
the time-evolution operator is given by
\begin{align}
U_k^\text{T}(t)
&= \ee^{-\ii t J_0(F) \cos k \sigma_x} \ee^{ -\ii z_k [\cos(\Omega t)-1]\sigma_x}\\
&= \ee^{-\ii [t J_0(F) \cos k -z_k]\sigma_x}\sum_n (-\ii \sigma_x)^n J_n(z_k)\ee^{\ii n\Omega t}\label{eq:UTFourier}
\end{align}
with
\begin{align}
z_k&\equiv \frac{2t_0J_1(F)}{\Omega}\sin k \sim \frac{t_0F}{\Omega}\sin k\label{eq:zk}.
\end{align}
Here the superscript T indicates that the truncation is performed,
and we have recovered the transfer integral $t_0$, which have been set to $1/2$ so far.
The contributions proportional to $J_n(F)$ with $|n|\ge 2$, which are $O(F^2)$,
are neglected in this approximation.

The solutions of the time-dependent Schr\"{o}dinger equation
is immediately obtained within this approximation
since $U_k^\text{T}(t)$ contains only $\sigma_x$.
By applying $U_k^\text{T}(t)$ onto
the eigenstates of $\sigma_x$,
$\sigma_x \sxe_\pm=\pm\sxe_\pm$ i.e.\  $\sxe_\pm=\begin{sumipmatrix}1\\ \pm1\end{sumipmatrix}/\sqrt{2}$,
we obtain from Eq.~\eqref{eq:UTFourier}
\begin{align}\label{eq:Tsol}
U^\text{T}_k(t) \sxe_\alpha = \ee^{\ii \alpha z_k -\ii E_\alpha(k)t} \sum_n (-\ii \alpha)^n J_n(z_k)\ee^{\ii n \Omega t} \sxe_\alpha
\end{align}
for $\alpha=\pm$,
where we have introduced
\begin{align}
E_\alpha(k) &=\alpha J_0(F)\cos k.
\end{align}
Comparing Eqs.~\eqref{eq:flsol} and \eqref{eq:Tsol},
one finds that
the two Floquet eigenvalues are 
$E_\pm(k)$
and their eigenstates are given by
\begin{align}\label{eq:FE0}
\chiv^\pm_n(k) =\frac{( \pm\ii)^n}{\sqrt{2}}J_n(z_k)\begin{pmatrix}1\\ \pm1 \end{pmatrix} \equiv c_n^\pm(k)\sxe_\pm.
\end{align}
Here we have ignored the phase factor $ \ee^{\pm\ii z_k}$
and this corresponds to the choice of the global phase of $\chiv^{\alpha}(k,t)$.
We note that we do not require $E_\alpha(k) \in [-\Omega/2,\Omega/2)$ in the analytical calculations for convenience.

The above approximation is equivalent to truncating the off-diagonal elements in the Floquet Hamiltonian as
\begin{align}\label{eq:HFT}
\hftmat_{mn}(k) \equiv
\begin{cases}
\hfmat_{mn}(k) & (|m-n|\le 1) \\
0 & (\text{otherwise}).
\end{cases}
\end{align}
The complete set of the eigenstates $\fechi^{\alpha,M}(k)$
of the truncated Floquet Hamiltonian $\hftmat(k)$
are defined for integer $M$'s as 
\begin{align}\label{eq:complete}
\chiv_n^{\alpha,M}(k) = \chiv^\alpha_{n+M}(k) = c_{n+M}^\alpha (k) \sxe_\alpha.
\end{align}
Then 
they satisfy the eigenvalue equation
\begin{align}\label{eq:eigen0}
\sum_n \hftmat_{mn} (k) \chiv^{\alpha,M}_n(k) = [E_\alpha(k)+M\Omega]\chiv^{\alpha,M}_m(k).
\end{align}

Now we discuss the distribution of a Floquet eigenstate over the Floquet space, or index $n$:
\begin{align}
\hhd_n^\alpha(k) \equiv |\chiv_n^\alpha(k)|^2 = J_n(z_k)^2.
\end{align}
As we have seen above, this distribution is approximately constant for $|n| \lesssim |z_k|$
and rapidly decays for $|n| \gtrsim |z_k|$.
Therefore, if $\Omega$ is smaller enough than $t_0$,
a plateau with width $|z_k|$ emerges in $\hhd_n^\alpha(k)$ even for $F<1$
where no plateau appears in the HHC.
In other words,
the width of the plateau in $\hhd_n^\alpha(k)$ is larger
than that in the HHC by the factor
\begin{align}
\enhance =\frac{\avz}{|F|}\sim \frac{t_0}{\Omega},
\end{align}
for $\Omega<t_0$, where
$\avz$ is the $k$-space average of $|z_k|$.

We note that,
in the single-band limit,
the plateau in $\hhd_n^\alpha(k)$ has nothing to do with the HHC spectrum.
This is because the time-dependent state~\eqref{eq:Tsol} is always proportional to
either of $\sxe_\alpha$'s and the high-harmonic terms in Eq.~\eqref{eq:Tsol} amount to an overall phase factor.
In fact, we have shown in Sec.~\ref{sec:intra} that
the HHC spectrum does not show a plateau for $|F|<1$,
although $\hhd_n^\alpha(k)$ can show a plateau.
We will show in Sec.~\ref{sec:pert}, however, that
the plateau in $\hhd_n^\alpha(k)$ is converted into the HHC spectrum
once the staggered potential is turned on.

We remark on the work in Ref.~\cite{Higuchi2014} that studied the case of quasistatic
input field with frequency $\Omega$ much smaller than the Bloch frequency $\ob$.
This corresponds to the limit of $F=\ob/\Omega\to\infty$ in the present study,
and the Bloch oscillation occurs many times within one period of input time dependence. 
Although this differs from the typical situation in the present study, one can apply the present
formulation without any problem also to this parameter regime, as far as the input field is periodic in time.
The harmonics in output are multiples of $\Omega$,
which distribute densely in the frequency space for small $\Omega$
and one of them is very close to the Bloch frequency, $\nb \Omega\sim \ob$.
Considering that the Floquet Hamiltonian has the largest matrix
element for this harmonics $\nb$,
we expect that the output spectrum shows
peaks around multiples of $\nb \Omega$,
which is consistent with the result of Ref.~\cite{Higuchi2014} predicting harmonics of $\ob$.
For this regime, one needs to diagonalize the Floquet Hamiltonian with a very large dimension greater than $\nb$,
and we do not further analyze this case.

\section{New Plateau induced by staggered potential}\label{sec:pert}
In this section,
we study the case of $Q\neq0$ and examine the effects of the staggered potential on the HHC spectrum.
We will develop an analytical perturbative approach to the effects of the staggered potential $Q$,
and mainly focus on the region of $|F|<1$ since the eigenstates~\eqref{eq:complete} are available.
While the $Q=0$ limit does not show a plateau,
we will show that a plateau appears in the HHC spectrum at the order of $Q^2$.

Since $|F|\lesssim 1$,
we may use the truncated Floquet Hamiltonian $\hftmat(k)$~\eqref{eq:HFT}.
We expand its eigenstates 
as a polynomial in $Q$:
\begin{align}\label{eq:fe_pert_exp}
\fechi^{\alpha,M}(k) +Q \fenu^{\alpha,M}(k) + Q^2 \felam^{\alpha,M}(k)+\cdots,
\end{align}
where $\fechi^{\alpha,M}(k)$ is the Floquet eigenstates in the single-band limit $(Q=0)$
discussed in Sec.~\ref{sec:fe_single}.
Correspondingly,
we expand the HHC spectrum for each Floquet eigenstate~\eqref{eq:hhceigen}
also as a polynomial in $Q$:
\begin{align}
I^{\alpha,0}_\text{H} (k,N)
+Q I^{\alpha,1}_\text{H} (k,N)
+Q^2 I^{\alpha,2}_\text{H} (k,N)+\cdots,
\end{align}
and we will calculate $I^{\alpha,1}_\text{H} (k,N)$ and $I^{\alpha,2}_\text{H} (k,N)$ below.

\subsection{Perturbation Theory}\label{sec:pert1}
The perturbation to the Floquet Hamiltonian is
\begin{align}\label{eq:VF}
\pert_{mn} = Q \sigma_z \delta_{mn}
\end{align}
and this interchanges the two eigenstates $\sxe_\pm$ of $\sigma_x$.
Its matrix elements between 
the unperturbed eigenstates are given by
\begin{align}\label{eq:Vmatel}
\fechi^{\alpha,M}(k)^\dag V^\text{F} \fechi^{\beta,M'}(k) = Q (\alpha\ii)^{M-M'}J_{M-M'}(2z_k) \delta_{\alpha,-\beta}.
\end{align}
Let us focus on the correction for $M=0$
owing to the physical equivalence of the Floquet eigenstates.
The first-order correction $\fenu^{\alpha,0}(k)$ is obtained by the standard first-order perturbation theory
as
\begin{align}\label{eq:firstp}
\fenu^{\alpha,0}(k)
=\sum_{M\neq0} \coeffF_M^\alpha(k) \fechi^{-\alpha,M}(k)
\end{align}
with
\begin{align}\label{eq:firstcoeff}
\coeffF_M^\alpha(k) = \frac{(-\alpha\ii)^M J_M(2z_k)}{M\Omega+2\alpha J_0(F)\cos k}.
\end{align}
We note that $|\coeffF_M^\alpha(k)|$ also shows a plateau structure in $M$
due to the Bessel function.
When the vanishing of the denominator of Eq.~\eqref{eq:firstp} is ignored,
the $M$-dependence of $|\coeffF_M^\alpha(k)|$ is approximately constant for $|M|\lesssim 2|z_k|$
and rapidly decays for larger $|M|$
as shown in Fig.~\ref{fig:i2}(a).

Although the correction $\fenu^{\alpha,0}(k)$ spreads over various Floquet bands,
the HHC does not change at the first order of $Q$,
or the first-order contribution to the HHC vanishes:
\begin{align}
\curH^{\alpha,1}(k,N)=0.
\end{align}
This is because
the current matrix~\eqref{eq:jmat} is proportional to $\sigma_x$
and $\fechi^{\alpha,0\dag}\sigma_x\fenu^{\alpha,M}=0$.

We make a remark on the vanishing of the denominator in Eq.~\eqref{eq:firstp} for some $k$.
This condition implies the resonance between two Floquet eigenstates
and one needs a degenerate perturbation analysis.
We show in Appendix~\ref{sec:resonance}
that this resonance actually gives contributions of $O(Q)$.
However, this contribution has the same $N$ dependence as the HHC spectrum in the single-band limit~\eqref{eq:Nth0k},
and, hence, does not show a plateau for $|F|\lesssim 1$.

Let us evaluate the $O(Q^2)$ correction of the HHC spectrum 
due to the matrix elements between $\fenu^{\alpha,M}$'s.
This is a part of $\curH^{\alpha,2}(k,N)$ and we define this as 
\begin{align}\label{eq:cur_corr2}
\curH^{\alpha,2A}(k,N)
\equiv \sum_{n,l} \nuv^{\alpha,0}_{n-l+N}(k)\jmat_{k,l}\nuv^{\alpha,0}_n(k).
\end{align}
In fact, another contribution of $O(Q^2)$ comes from the second-order correction of
the wave function $\felam^{\alpha,M}(k)$.
In Appendix~\ref{sec:pert2},
we show that its $N$-dependence is similar to that of Eq.~\eqref{eq:cur_corr2}.
By invoking Eqs.~\eqref{eq:firstp}
and performing some algebra,
we obtain
\begin{align}
&\curH^{\alpha,2A}(k,N)\notag\\
&=\alpha e\sum_{M,M'} \coeffF_M^\alpha(k)^* \coeffF_{M'}^\alpha(k) \curcoeff_{M-M'+N}(k).\label{eq:hhc2nd}
\end{align}

\bfig
\includegraphics[width=8cm]{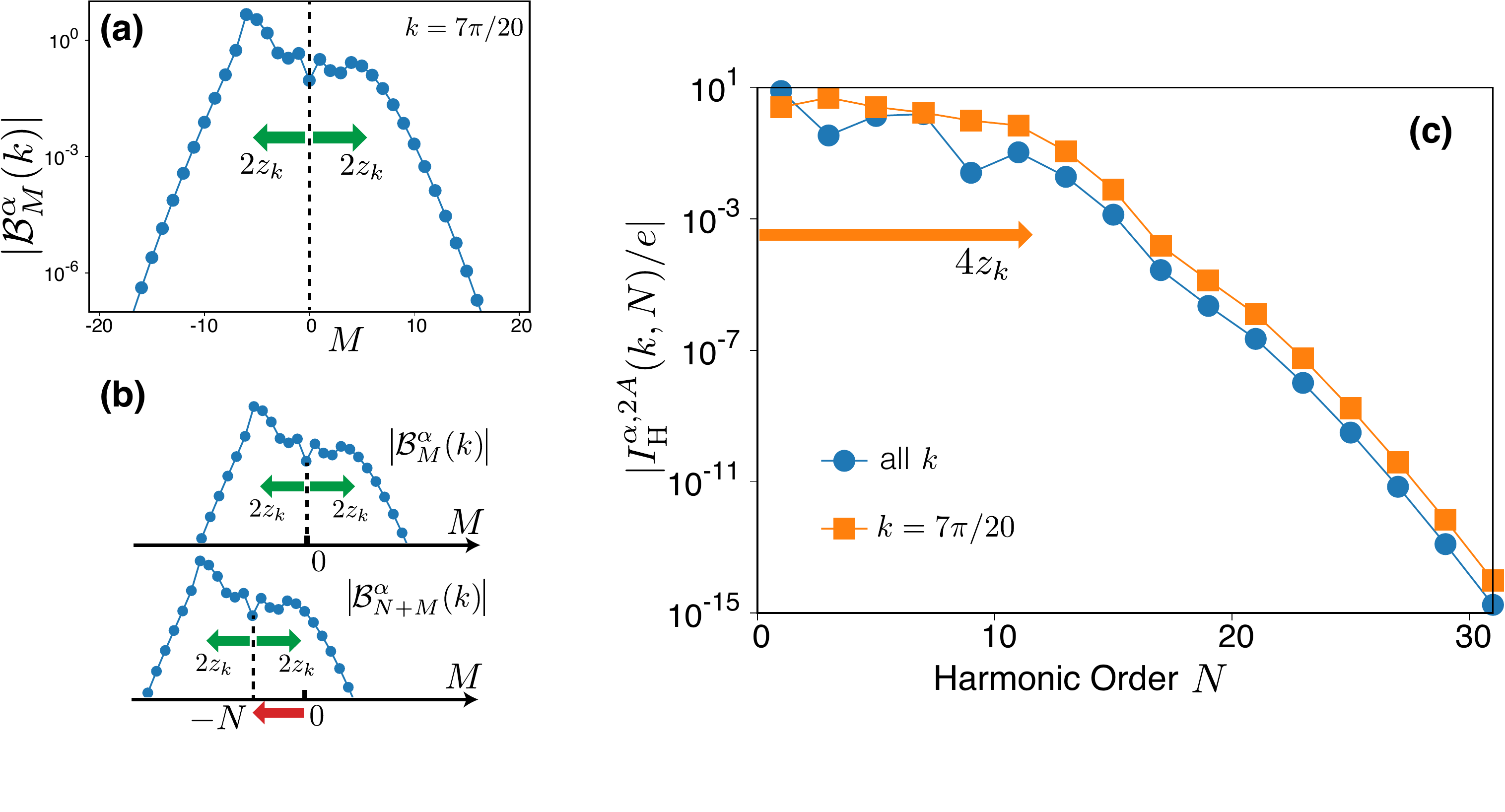}
\caption{
(a) Typical behavior of $|\coeffF_M^\alpha(k)|$ [Eq.~\eqref{eq:firstcoeff}] calculated for $F=0.5$, $\Omega=0.15$, $\alpha=+$, and $k=7\pi/20$.
The plateau region $M\lesssim 2z_k$ is shown by arrows.
(b) Schematic illustration for the overlap between $\coeffF_M^\alpha(k)$ and $\coeffF_{N+M}^\alpha(k)$.
For $N\gtrsim 4z_k$, the overlap rapidly decays.
(c) The second-order correction $\curH^{\alpha,2}(k,N)$ to the HHC spectrum~\eqref{eq:cur_corr2}
for $k=7\pi/20$ (square) and for the average over $k=(\pi/2)(m/10)$ with $m=0,1,\dots,9$.
The arrow indicates $4z_k$ for $k=7\pi/20$.
}
\label{fig:i2}
\efig

Equation~\eqref{eq:hhc2nd} implies that, when $4|z_k| >1$,
the HHC spectrum $I_\text{H}^{\alpha,2A}(k,N)$ shows a plateau for $|N|\lesssim 4|z_k|$
and rapidly decays for larger $|N|$
as understood as follows.
Since we are considering $|F|\lesssim1$
and $\curcoeff_n(k) \propto J_n(F)$ rapidly decays as $|n|$ increases,
the sum over $M'$ is dominated by $M'= N+M$ and
Eq.~\eqref{eq:hhc2nd} is approximated as
$\alpha e v_0(k)\sum_{M}\coeffF^\alpha_M(k)^*\coeffF^\alpha_{N+M}(k)$.
As Fig.~\ref{fig:i2}(b) shows,
this sum, or the overlap between $\coeffF^\alpha_M(k)$ and $\coeffF^\alpha_{N+M}(k)$,
rapidly decays for $|N|\gtrsim 4|z_k|$,
whereas it changes rather slowly for $|N|\lesssim 4|z_k|$~\footnote{Near $k=0$, rapid decays actually occur for $|N|\lesssim 4|z_k|$.
This is because $M\Omega$ plays a minor role in Eq.~\eqref{eq:firstcoeff} compared with $2\alpha J_0(F)\cos k$.
In fact, if we neglect $M\Omega$, 
we have $\curH^{\alpha,2}(k,N)=-\alpha [2J_0(F)\cos k]^{-2}v_{N}(k)\propto J_N(F)$,
which rapidly decays with $N$.
Thus the plateau of the HHC spectrum is contributed from larger $k$ values.
}.
This is how a plateau appears in the HHC spectrum
at the second order of the staggered potential $Q$.
The above argument is confirmed by the numerical results shown
in Fig.~\ref{fig:i2}(c),
where $\curH^{\alpha,2A}(k,N)$ is calculated as in Eq.~\eqref{eq:cur_corr2} and plotted
for a representative $k$ and the average over $k$.

As a result of these analyses,
we propose 
\begin{align}\label{eq:ncut}
N_\text{cut} = \frac{8}{\pi} |J_1(F)|\frac{t_0}{\Omega}
\end{align}
as an indicator of the plateau width, or the high-energy cutoff order of the HHC spectrum.
Here we have recovered the transfer integral $t_0$,
which has been set to $1/2$.
The observable of interest is actually
$\sum_k \curH^{\alpha,2A}(k,N)$
and $4z_k$ depends on $k$.
Averaging $|z_k|$ over the Brillouin zone,
we obtain Eq.~\eqref{eq:ncut}.
Since the averaging is a crude approximation,
the numerical factor $8/\pi$ in Eq.~\eqref{eq:ncut} should not be taken very seriously.

Equation~\eqref{eq:ncut} can be used to derive the onset field strength $F_\text{onset}$ 
at which the plateau sets in.
Whether a plateau exists or not
should correspond to $N_\text{cut}\lesssim1$ and $N_\text{cut}\gtrsim1$, respectively.
Thus $F_\text{onset}$ is estimated by the condition $N_\text{cut}=1$,
which leads to
\begin{align}\label{eq:Fonset}
F_\text{onset} = \frac{\Omega}{t_0},
\end{align}
where we have ignored the numerical factor $\pi/4$ and approximated $J_1(F)$ by $F/2$ assuming $F_\text{onset}$ is small enough.
We emphasize that $F_\text{onset}$ can be less than 1
if $\Omega<t_0$.

\subsection{Scaling properties of the plateau}
Now we discuss how $N_\text{cut}$ depends on the amplitude $E_0$
and the frequency $\Omega$
of the input ac electric field.
When $|F|\lesssim 1$,
we approximate $J_1(F)\simeq F/2$ in Eq.~\eqref{eq:ncut}
and omit the numerical factor to obtain
\begin{align}\label{eq:ncut_p}
N_\text{cut}  \sim \frac{|eaE_0 t_0|}{(\hbar\Omega)^2},
\end{align}
where we have used Eq.~\eqref{eq:Fdef} and recovered $a$ and $\hbar$ that have been set to unity.

Equation~\eqref{eq:ncut_p} shows that 
the cutoff order $N_\text{cut}$ is proportional to the amplitude $E_0$
rather than the power $E_0^2$ of the input electric field.
This is consistent with the experimental observations~\cite{Ghimire2011} and
a unique feature of the HHG in solids in contrast to that in gases.

A remarkable prediction of
Eq.~\eqref{eq:ncut_p} is that the cutoff order $N_\text{cut}$ 
is proportional to $\Omega^{-2}$ rather than $\Omega^{-1}$ for a fixed field amplitude $|E_0|$.
This originates from the intrinsic property of the Floquet eigenstate $\fechi^{\alpha,M}(k)$.
As shown in Sec.~\ref{sec:fe_single}, this eigenstate distributes over the Floquet index
and the width of the distribution is proportional to $\avz\sim J_1(F)/\Omega \sim F /\Omega\propto \Omega^{-2}$
since $F=\Omega_\text{B}/\Omega$.
Thus the scaling $N_\text{cut}\propto \Omega^{-2}$ is a signature of the Floquet eigenstate,
which could be tested in experiments.

We remark that
the cutoff energy defined by
\begin{align}
E_\text{cut} \equiv N_\text{cut}\hbar\Omega \sim \frac{|eaE_0 t_0|}{\hbar\Omega}
\end{align}
has a slightly different scaling.
The cutoff energy is proportional to $E_0$ and $\Omega^{-1}$.
This scaling could also be tested experimentally if several frequencies for the input are available.

\subsection{Numerical Verification}
Let us numerically verify the above analytical arguments based on perturbation theory.
Figure~\ref{fig:pert_plateau} shows the HHC spectrum $\curH (N)$ calculated as in Eq.~\eqref{eq:defhhc}
for several parameter sets $(F,\Omega)$ with $L=10^4$ and $Q=0.01$.
We have used 
the Floquet Hamiltonian without truncation $\hfmat_{mn}(k)$,
and the cutoff for the Floquet index has been chosen as $-80\le m,n\le 80$.

\bfig
\includegraphics[width=8cm]{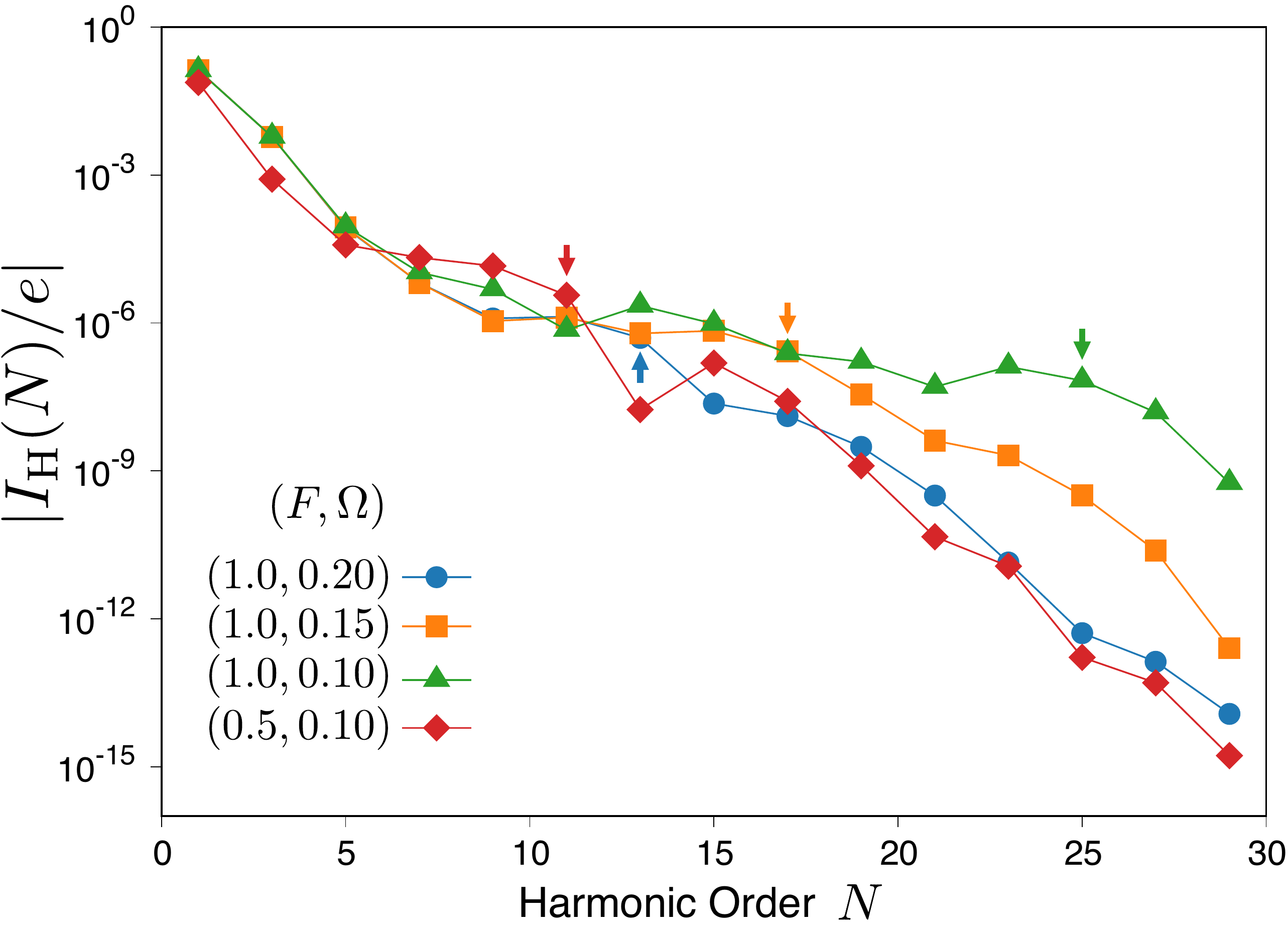}
\caption{
The HHC spectrum calculated for the four parameter sets $(F,\Omega)$
as indicated in the figure.
The parameter $Q$ is set $0.01$.
The arrow indicates the end of the plateau that is identified from the plot.
}
\label{fig:pert_plateau}
\efig

A plateau is observed for each parameter set $(F,\Omega)$ in Fig.~\ref{fig:pert_plateau}.
For the $F=1.0$ data,
we assign from the plot the cutoff order $N_\text{cut}$ as 13 ($\Omega=0.2$), 17 ($\Omega=0.15$), and 25 ($\Omega=0.1$)
as indicated by arrows in the figure.
The ratios between
these numbers are in good agreement with the analytical prediction~\eqref{eq:ncut},
which states that $N_\text{cut}$ is proportional to $\Omega^{-1}$ with fixed $F$.
We also assign $N_\text{cut}=11$ for the $(F,\Omega)=(0.5,0.1)$ data,
which is approximately consistent with Eq.~\eqref{eq:ncut}.

We have also verified that
the $Q$-dependence of the magnitude $|\hhc{N}|$
is consistent with the perturbation analysis.
This requires 
a careful treatment due to the resonances between the Floquet eigenstates
as remarked in Sec.~\ref{sec:pert1}.
We show the details of the verification in Appendix~\ref{sec:resonance}.

\section{Beyond Perturbation}\label{sec:beyond}
In Secs.~\ref{sec:single} and \ref{sec:pert}, we have analytically investigated the HHC spectrum in the two limiting cases:
(i) $Q=0$ and arbitrary $F$, and (ii) $|F|\lesssim 1$ and small $|Q|$.
In the other cases, we numerically calculate the HHC spectrum
and show that the scalings in the limiting cases still hold
if either $|F|$ or $|Q|$ is small, whereas a qualitative difference sets in when both $|F|$ and $|Q|$ become large.

\begin{figure*}
\includegraphics[width=17cm]{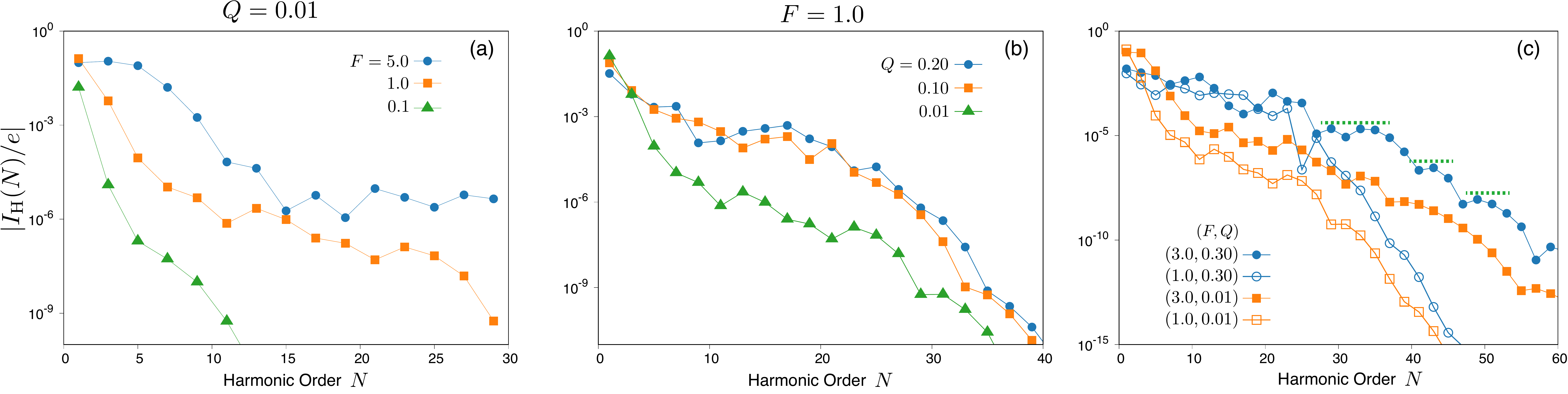}
\caption{
Numerically calculated HHC spectrum in the nonperturbative regimes
including the three cases (a) small $Q$ but large $F$, (b) small $F$ but large $Q$, and (c) large $F$ and $Q$.
In all panels, $\Omega=0.1$.
In panel (a), the three data sets correspond to different field strengths $F=5.0$ (circle), $1.0$ (square), and $0.1$ (triangle)
with $Q=0.01$.
In panel (b),
the staggered potential strength is varied: $Q=0.20$ (circle), 0.10 (square), and 0.01 (triangle) with $F=1.0$.
In panel (c), the parameter set is $(F,Q)=$
(3.0,0.30) (filled circle), (1.0,0.30) (open circle), (3.0,0.01) (filled square), and (1.0,0.01) (open square).
Dashed lines indicate the multi-step plateaus.
}
\label{fig:mp}
\end{figure*}

First, we investigate the case of $|F|\ge1$ with very small $|Q|$.
Figure~\ref{fig:mp}(a) shows the HHC spectrum for $F=1.0$ and 5.0
with $\Omega=0.1$ and $Q=0.01$.
For $F=5.0$,
we observe two plateaus in $1\le N\le 5$ and $N\ge11$, respectively.
The first plateau originates 
from the contribution discussed in Sec.~\ref{sec:single}
since its width coincides with the value of $F$.
The second plateau already exists at $F=1.0$, and it is induced by the staggered potential.
We note that the width of the second plateau does not necessarily follow Eq.~\eqref{eq:ncut}
since the truncation of the Floquet Hamiltonian is no longer justified for $F\gtrsim 1$.

Second, we discuss the case of $|F|\lesssim1$ with larger $|Q|$.
Figure~\ref{fig:mp}(b) shows
the HHC spectrums for $Q=0.1$ and $0.2$ with $(F,\Omega)=(1.0,0.1)$.
They are compared with the data for $Q=0.01$, for which the perturbation analysis works well.
For $Q=0.1$ and $0.2$,
the perturbation theory 
is no longer justified
since $Q$ is not the smallest parameter,
but its results are still valid approximately.
Namely, 
compared with the data for $Q=0.01$,
the plateau has almost the same width,
and its height is enhanced about two orders of magnitude.
Thus we conclude that the presence of the plateau induced by the staggered potential
is not restricted to the region $|Q| <\Omega$, but can be extended to $|Q|>\Omega$
for $|F|\lesssim1$.

Finally, we consider the case that neither $|F|$ nor $|Q|$ is small.
Figure~\ref{fig:mp}(c) shows that the HHC spectrum for $Q=0.3$ and $F=3.0$
is remarkably enhanced for $N\gtrsim 30$ compared with the data for $F=1.0$.
Moreover, in addition to the plateau in $N\lesssim 20$,
multi-step plateaus emerge as indicated by dashed lines in the figure.
We also plot the data for $Q=0.01$ in Fig.~\ref{fig:mp}(c) for comparison.
Although the enhancement of the magnitude with $F$
is common for both $Q$'s,
the multi-step plateaus appear only for the larger $Q$.
Thus a qualitative difference arises in the HHC spectrum when neither $|F|$ nor $|Q|$ is small,
and an approach beyond perturbation is desired to reveal the origin of the multi-step plateaus.

We make a remark on the possible relevance of the plateaus in $N\gtrsim30$ in experiments.
Ndabashimiye and coworkers~\cite{Ndabashimiye2016} have observed the HHG in rare-gas solids
and reported that a new plateau emerges around $N\sim 30$
as the input laser intensity is increased.
This experimental observation is consistent at least apparently with our numerical results.
Thus the plateaus in Fig.~\ref{fig:mp}(c) merit further study.

\section{Conclusions}\label{sec:conclusions}
We have considered the setup where an ac electric field with frequency $\Omega$ is turned on at some time.
In this setup, the harmonics are well defined as multiples of $\Omega$,
and the HHC spectrum have been related to the Floquet eigenstates [see Eq.~\eqref{eq:defhhc} and \eqref{eq:hhceigen}].
Our formulation is a generalization of similar formulas in the literature
because the condition that the initial state is the ground state
is taken into account by the weights of each Floquet eigenstate.
In this formulation,
analytical approaches are feasible,
and the consequences of symmetries of the Hamiltonian are easily tractable
from the Floquet eigenstates and their weights.

On the basis of this formulation, we have investigated
the HHC spectrum of electrons on a one-dimensional chain
with the staggered potential $Q$,
which splits the single cosine band into two.
In the single-band limit ($Q=0$), we have confirmed the result~\cite{Pronin1994}
that a plateau of width $|F|$ appears
owing to the nonlinearity of the Peierls substitution for strong field $|F|>1$,
where the dimensionless parameter $F=\ob/\Omega\propto E_0/\Omega$ quantifies the coupling between
the electron and the electric field.
Our new finding is that
the staggered potential $Q$ induces another wider plateau emerging from weaker field $|F|<1$.
On the basis of the asymptotically exact solutions of the Floquet eigenstates for $Q=0$,
we have shown that
the width of plateau induced by the staggered potential is $\enhance |F|=|eaE_0t_0|/(\hbar\Omega)^2$, which
is proportional to $E_0$ and larger by the factor $\enhance=t_0/\Omega$ than that in the absence of the staggered potential.
Our result also provides a new prediction that the width of the plateau scales as $\Omega^{-2}$.
Since this scaling originates from the Floquet eigenstates,
it could also be an experimental signature in identifying those states.

We have numerically confirmed that our analytical results hold qualitatively 
as far as either the field amplitude $|F|$ or the staggered potential $|Q|$ is small.
This condition includes the case where $\Omega$ is smaller than the band gap.
We have also analyzed the case where both $|F|$ and $|Q|$ are large
and found that a more complex structure in the HHC spectrum involving the multi-step plateaus,
which might be relevant in interpreting experiments and merits further systematic studies.

We make a remark on an implication on the HHG experiments in CDW materials.
The cases $Q=0$ and $Q\neq0$ correspond to the phases above and below
the transition temperature $\tc$ of the CDW order.
Because of high carrier density above $\tc$,
the largest $|E_0|$ is limited in experiments to avoid damaging the samples
and only a few harmonics would be observable.
In fact, the 9th harmonic has been the highest observed in 2H-NbSe${}_2$ above $\tc$~\footnote{
K. Shimomura, K. Uchida, K. Nagai, and K. Tanaka (unpublished)}.
Our results imply that, even with the same limited $|E_0|$,
the HHG is enhanced below $\tc$ and a plateau could be observable in the spectrum. 
In this situation, our results also predict that the cutoff order $N_\text{cut}$
scales as $\Omega^{-1}$ and $\Omega^{-2}$ above and below $\tc$, respectively.

As a concluding remark,
we should mention that several physical processes are not taken into account in our formulation.
First, the correlation effects~\cite{Ikemachi2017a,Silva2018,Murakami2017,Murakami2018}
and the order parameters~\cite{Nag2018} have been neglected in our model.
Second, the energy dissipation to the phonon thermal bath~\cite{Hamilton2015}. has not been considered either,
which might be relevant if the driving frequency is as low as 1THz$\sim \mathrm{1ps}^{-1}$.
For the carrier bath, the Keldysh formalism has been employed to calculate the HHC~\cite{Zhou2011,Morimoto2016,Morimoto2016a,Kim2017}.
Third, we have treated the electric field classically.
Quantum processes lead to spontaneous emission of photons~\cite{Yamane2018a},
which is not discussed in the present work.
These effects are all important
and our simple model could serve as a starting point
in interpreting the experimental data.

\section*{Acknowledgements}
Fruitful discussions with Y. Kayanuma, K. Shimomura, T. Tamaya, and K. Uchida
are gratefully acknowledged.
This work was supported by
JSPS KAKENHI Grants No.~JP16H06718 and JP18K13495.
K.C. acknowledges financial support
provided by the Advanced Leading Graduate Course for
Photon Science at the University of Tokyo.

\appendix
\section{Resonances of Floquet eigenstates}\label{sec:resonance}
In this appendix,
we refine the perturbation theory in Sec.~\ref{sec:pert1}
by taking account of the resonances of the Floquet eigenstates,
and calculate the $O(Q)$ contributions to the HHC spectrum.

Let us take a positive integer $\res$
satisfying $\res < |2J_0(F)/\Omega|$.
Then there exist a momentum $\resmom$ such that
\begin{align}
E_+(\resmom)-\res\Omega = E_-(\resmom).
\end{align}
This means that when $Q=0$ the Floquet eigenstates $\fechi^{+,\res}(\resmom)$ and $\fechi^{-,0}(\resmom)$ are degenerate.
We investigate the Floquet eigenstates for $Q\neq0$ in the vicinity of $k=\resmom$,
where the mixing of these states cannot be treated by perturbation theory with respect to $Q$,
and must be treated exactly.

Since we are interested in the vicinity of $k=\resmom$,
the $2\times2$ Hamiltonian matrix
within the subspace spanned by $\fechi^{+,\res}(\resmom)$ and $\fechi^{-,0}(\resmom)$
is linearized in $\Delta k \equiv k-\resmom$,
and we obtain
\begin{widetext}
\begin{align}\label{eq:mat2b}
\hres(k)=
\begin{pmatrix}
E_+(k)-\res\Omega  &  Q\ii^{\res} J_{\res}(2z_k) \\
Q(-\ii)^{\res} J_{\res}(2z_k)& E_-(k)
\end{pmatrix}
\sim E_-(\resmom)\idtwo +
\begin{pmatrix}
-a & \ii^\res b \\ (-\ii)^\res b  & a 
\end{pmatrix},
\end{align}
\end{widetext}
where $\idtwo$ is the unit matrix, $z_k=2J_1(F)\sin k/\Omega$,
and we have defined
$a \equiv \velres_\res\Delta k$,
$b \equiv (c_\res+d_\res\Delta k)Q$,
$\velres_\res \equiv J_0(F)\sin \resmom$,
$c_\res \equiv J_\res(2z_{\resmom})$,
and 
$d_\res \equiv 2J_\res'(2z_{\resmom}) \res J_1(F)/J_0(F)$.

The two eigenvalues of the linearized Hamiltonian~\eqref{eq:mat2b} are given by
\begin{align}\label{eq:eval_split}
E_-(\resmom) \pm \Delta E(\Delta k)
\end{align}
with the energy splitting
\begin{align}\label{eq:eval_diff}
\Delta E( \Delta k) \equiv \sqrt{ Q^2(c_\res+d_\res\Delta k)^2+\velres_\res^2(\Delta k)^2 },
\end{align}
and the two-fold degeneracy is lifted by the coupling $Q$.
We emphasize that the energy splitting is not minimal at $\Delta k=0$ in general.
The position of minimum, which is denoted by $\Delta k_*$, is obtained by
minimizing Eq.~\eqref{eq:eval_diff} as
\begin{align}
\Delta k_* = -\frac{Q^2 c_\res d_\res}{\velres_\res^2+Q^2d_\res^2}.
\end{align}
The position of resonance shifts by this amount due to the coupling $Q$.

We denote by $^\text{t}\!\!\begin{pmatrix} x_\pm(k) , y_\pm(k) \end{pmatrix}$
the two eigenvectors with the corresponding eigenvalues~\eqref{eq:eval_split}.
The explicit forms of these eigenvectors are given by
\begin{widetext}
\begin{align}
\begin{pmatrix} x_\pm(k) \\ y_\pm(k) \end{pmatrix}
= \frac{1}{\sqrt{2\Delta E(\Delta k)}[\Delta E(\Delta k)\mp a]^{1/2}} 
\begin{pmatrix} \pm\Delta E(\Delta k)-a \\ (-\ii)^\res b\end{pmatrix}.
\end{align}
\end{widetext}
Thus the appropriate Floquet eigenstates are given by
\begin{align}\label{eq:flnew}
\fechinew^+(k) & = x_+(k) \fechi^{+,\res}(k) + y_+(k) \fechi^{-,0}(k),\\
\fechinew^-(k)  &= x_-(k) \fechi^{+,\res}(k) + y_-(k) \fechi^{-,0}(k)
\end{align}
instead of $\fechi^{+,\res}(k)$ and $\fechi^{-,0}(k)$ in the vicinity of resonance.

The new Floquet eigenstates $\fechinew^\pm(k)$ carries harmonic currents,
which are proportional to $\Delta k$.
From Eqs.~\eqref{eq:jmat}, \eqref{eq:defhhc}, and \eqref{eq:hhceigen}, we obtain
\begin{align}
I_\text{H}^{\text{res}} (k,N)
&\equiv -e\curcoeff_N(k)\sum_{\alpha=\pm}|\flcoeff_\alpha(k)|^2 \left[ |x_\alpha(k)|^2-|y_\alpha(k)|^2\right]\\
&= -\frac{e\curcoeff_N(k) \velres_\res \Delta k}{\Delta E(\Delta k)}\left[ |\flcoeff_-(k)|^2-|\flcoeff_+(k)|^2\right],
\end{align}
where $\flcoeff_\pm(k)$ are now expansion coefficients in terms of $\fechinew^\pm(k)$.
This result reflects the fact that the Floquet eigenstates $\fechi^{+,\res}(\resmom)$ and $\fechi^{-,0}(\resmom)$
carry harmonic currents with opposite sign.

The HHC contribution near resonance $I_\text{H}^{\text{res}} (k,N)$
becomes as large as $O(Q^0)$ if $|\velres_\res \Delta k| \lesssim Q$,
although it vanishes at an exceptional point $\Delta k=0$,
where $|x_\pm(k)|^2 = |y_\pm(k)|^2=1/2$.
Thus, when summed over $k$, the HHC contribution near resonance 
amounts to $O(Q)$ owing to the $k$-space volume factor of $Q$.
We note that the $k$ sum around $\Delta k$ does not vanish
because
$\Delta E(\Delta k)$ is not an even function of $\Delta k$
and the denominator becomes minimum at $\Delta k=\Delta k_*\neq0$.

The $O(Q)$ contribution from resonance does not show a plateau for $|F|\lesssim1$
since it originates from the single-band limit discussed in Sec.~\ref{sec:single}.
In fact, the $N$-dependence of $I_\text{H}^{\text{res}} (k,N)$
derives from that of $\curcoeff_N(k)$ and hence $J_N(F)$.
As mentioned in Sec.~\ref{sec:single},
this does not show a plateau for $|F|\lesssim1$.

The resonance also has higher-order contributions of $O(Q^2)$.
Mixing of $\fechinew^\pm(k)$ with 
the other Floquet eigenstates with eigenvalues differring by $M\Omega$ ($M\in \mathbb{Z}$)
is caused by $Q$ at the first order,
and this mixing leads to $O(Q)$ contribution to the HHC spectrum.
When summed over $k$, this contribution amounts to $O(Q^2)$ due to the $k$-space volume factor of $Q$.
One can show that the $N$-dependence of this contribution can be a plateau,
but its width is about $2z_k$ rather than $4z_k$ obtained in Sec.~\ref{sec:pert1}.
Thus the $O(Q^2)$ contribution from resonances is not very important
to determine the width of the plateau in the HHC spectrum.

Let us now numerically verify the $Q$-dependence of the HHC spectrum for small $Q$.
For this purpose, we define the HHC spectrum induced by the staggered potential
\begin{align}\label{eq:deltaIH}
\Delta \hhc{N} \equiv \hhc{N} - \hhc{N,Q=0},
\end{align}
where $\hhc{N,Q=0}$ denotes the HHC spectrum in the single-band limit discussed in Sec.~\ref{sec:single}.
According to our perturbation theory,
this quantity is expanded as a polynomial in $Q$ as
\begin{align}\label{eq:ff}
\Delta \hhc{N} = e(a_NQ + b_N Q^2 +\cdots).
\end{align}
We have shown that
$|a_N|$ decreases faster than $|b_N|$
and the $O(Q^2)$ contribution becomes more important than the $O(Q)$ one for larger $N$.
Figure~\ref{fig:qscaling}(a) shows the numerically calculated $\Delta\hhc{N}$ for several $Q$'s
with $F=0.5$ and $\Omega=1.0$.
The $Q$-dependence of $\Delta\hhc{N}$ is fitted well by a polynomial~\eqref{eq:ff} of the fourth order,
and this justifies the polynomial expansion.
The first- and second-order coefficients $a_N$ and $b_N$ are shown in Fig.~\ref{fig:qscaling}(b).
This figure shows that the $O(Q)$ contribution decreases with $N$ faster than the $O(Q^2)$ one,
and this tendency is consistent with our analytical calculations.
Thus these numerical data support our analysis with perturbation in $Q$
including the resonance effects between the Floquet eigenstates.

\bfig
\includegraphics[width=8cm]{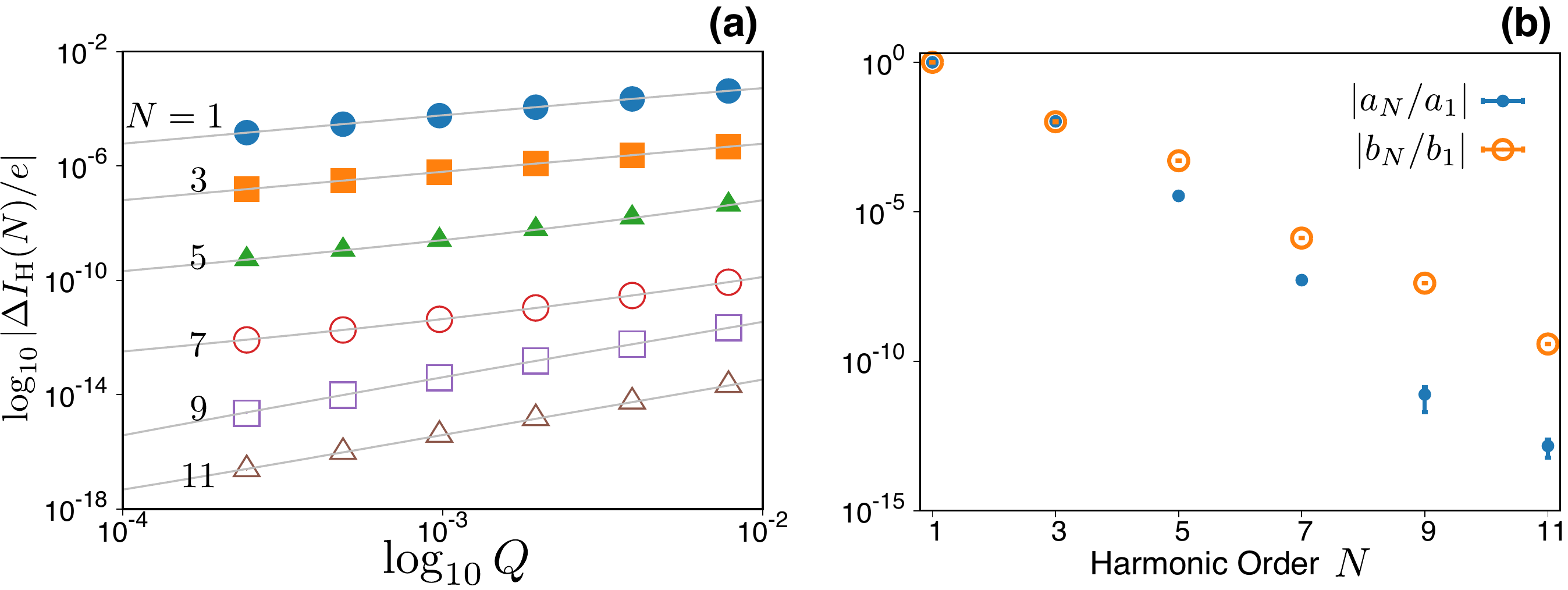}
\caption{
(a) Absolute value of $\Delta I_\text{H}(N)$ [Eq.~\eqref{eq:deltaIH}] calculated for $F=0.5$ and $\Omega=1.0$
plotted against $Q$ in the log-log scale.
Data are shown for the harmonics $N=$ 1 (filled circle), 3 (filled square),
5 (filled triangle), 7 (open circle), 9 (open square), and 11 (open triangle).
The solid lines are the polynomial fit~\eqref{eq:ff} of the fourth order.
(b) The determined fitting parameters normalized as $a_N/a_1$ (filled) and $b_N/b_1$ (open) plotted against the harmonic order $N$,
where $a_1=6.0\times10^{-2}$ and $b_1=-1.0$. The error bars show the fitting errors.
}
\label{fig:qscaling}
\efig

\section{Second-Order Perturbation Theory}\label{sec:pert2}
Here we extend the first-order perturbation theory in Sec.~\ref{sec:pert1}
to the second order.
We derive the second order correction $\felam^{\alpha,0}$
to the Floquet eigenstates,
and show that its $O(Q^2)$ contribution to the HHC 
has a similar $N$-dependence to the one discussed in Sec.~\ref{sec:pert1}.

The second-order correction $\felam^{\alpha,0}(k)$ in Eq.~\eqref{eq:fe_pert_exp}
is a superposition of $\fechi^{\alpha,M}(k)$ with various $M$'s
with the same $\sigma_x$-eigenvalue $\alpha$
since each $\pert_{mn}$ [Eq.~\eqref{eq:VF}] flips $\alpha$.
Thus the correction is represented as 
\begin{align}\label{eq:secondp}
\felam^{\alpha,0}(k) = \sum_{M\neq0} \coeff_M^\alpha(k) \fechi^{\alpha,M}(k),
\end{align}
and the coefficients $\coeff_M^\alpha(k)$ are given by the standard procedure
from the matrix elements~\eqref{eq:Vmatel} and the eigenenergies as
\begin{align}
\coeff_M^\alpha(k)=
\frac{(\alpha\ii)^M }{M\Omega}\sum_{M'} \frac{J_{M-M'}(2z_k) J_{-M'}(2z_k)}{M'\Omega + 2\alpha J_0(F)\cos k }.\label{eq:coeffs}
\end{align}
One can easily check that $\coeff_M^\alpha(-k)=(-1)^M \coeff_M^\alpha(k)=\coeff^{-\alpha}_{-M}(k)$.

We briefly interpret the $M$-dependence of $\coeff_M^\alpha(k)$ in Eq.~\eqref{eq:coeffs}.
We can safely ignore the vanishing of the denominator,
which corresponds to the resonance discussed in Appendix~\ref{sec:resonance},
because the contribution of the resonant region amounts to $O(Q^3)$ due to the
extra factor $Q$ from the $k$-space volume.
The sum in Eq.~\eqref{eq:coeffs} takes
the form of $\sum_{M'} g_{M'} J_{M-M'}(2z_k) J_{-M'}(2z_k)$
with a gradually changing $g_{M'}$.
Now we recall that the $n$-dependence of $|J_{n}(2z_k)|$ is approximately constant
for $|n|\lesssim 2z_k$ and decays rapidly for $|n|\gtrsim 2z_k$.
Therefore, the sum and, hence, $\coeff_M^\alpha(k)$ depend on $M$ rather slowly for $|M|\lesssim |4z_k|$
and rapidly decreases for $|M|\gtrsim |4z_k|$.
We note that this behavior is not qualitatively modified by
the overall factor $M^{-1}$ in Eq.~\eqref{eq:coeffs}
since it varies slowly.

Let us evaluate the second-order correction of the HHC spectrum
from $\felam^{\alpha,0}$:
\begin{align}
\curH^{\alpha,2B}(k,N)
&\equiv \sum_{n,l} \left(\chiv^{\alpha,0}_{n-l+N}(k)^\dag \jmat_{k,l} \lamv^{\alpha,0}_n(k)\right. \notag\\
&\qquad\quad\left.+\lamv^{\alpha,0}_{n-l+N}(k)^\dag \jmat_{k,l} \chiv^{\alpha,0}_n(k)\right).
\end{align}
By invoking Eq.~\eqref{eq:secondp}
and performing some algebra,
we obtain
\begin{align}
\curH^{\alpha,2B}(k,N)
&=-\alpha e\sum_{M\neq0} \left[ \curcoeff_{N-M}(k) \coeff_M^\alpha(k) + \curcoeff_{N-M}^*(k)\coeff_M^{-\alpha}(k)^*\right].
\label{eq:hhc2ndApp}
\end{align}
Here $\curcoeff_n(k)\propto J_n(F)$ has a significant weight only around $n=0$ since $|F|\lesssim1$.
Then the sum over $M$ in Eq.~\eqref{eq:hhc2nd} 
leaves 
$\alpha v_0(k) [ \coeff_N^\alpha(k) +\coeff_N^{-\alpha}(k)^*]$
and 
the $N$-dependence of $\curH^{\alpha,2B}(k,N)$ is governed by that of $\coeff^\pm_{N}(k)$.
Since $\coeff^\pm_{N}(k)$ shows a plateau as shown above, 
$\curH^{\alpha,2B}(k,N)$ also shows a plateau in its $N$-dependence.
The width of the plateau is $4|z_k|$.

We remark that, apart from the resonance, the whole $O(Q^2)$ contribution to the HHC spectrum
is given by the sum $\curH^{\alpha,2}(k,N)=\curH^{\alpha,2A}(k,N)+\curH^{\alpha,2B}(k,N)$,
where $\curH^{\alpha,2A}(k,N)$ derives from the first-order corrections to the Floquet eigenstates
explained in Sec.~\ref{sec:pert1}
and shows a plateau of width $4|z_k|$.
The analysis in this appendix shows that a similar plateau appears in the other part $\curH^{\alpha,2B}(k,N)$
with the same width.

\end{document}